\documentclass[11pt,a4paper]{article}
\pdfoutput=1 % if your are submitting a pdflatex (i.e. if you have
\usepackage[utf8]{inputenc}
\usepackage{jheppub}
\usepackage{graphicx}
\usepackage{caption}
\usepackage{subcaption}
\usepackage{hyperref}
\usepackage[T1]{fontenc}
\usepackage{makecell}

\newcommand{\be}{\begin{equation}}
\newcommand{\ee}{\end{equation}}
\newcommand{\bea}{\begin{eqnarray}}
\newcommand{\eea}{\end{eqnarray}}

\newcommand{\nn}{\nonumber\\}
\def\CN{\mathcal{N}}

\def\CO{\mathcal{O}}

\def\qfr{\mathfrak{q}}
\def\wfr{\mathfrak{w}}

\def\q{{\bf q}}

\title{Reconstruction of spectra and an algorithm based on the theorems of Darboux and Puiseux}

\author[a,b]{Sa\v{s}o Grozdanov}
\author[b]{and Timotej Lemut}
\affiliation[a]{Higgs Centre for Theoretical Physics, University of Edinburgh, Edinburgh, EH8 9YL, Scotland}
\affiliation[b]{Faculty of Mathematics and Physics, University of Ljubljana, Jadranska ulica 19, SI-1000 Ljubljana, Slovenia}

\abstract{
Assuming only a known dispersion relation of a single mode in the spectrum of a meromorphic two-point function (in the complex frequency plane at fixed wavevector) in some quantum field theory, we investigate when and how the reconstruction of the complete spectrum of physical excitations is possible. In particular, we develop a constructive algorithm based on the theorems of Darboux and Puiseux that allows for such a reconstruction of all modes connected by level-crossings. For concreteness, we focus on theories in which the known mode is a gapless excitation described by the hydrodynamic gradient expansion, known at least to some (preferably high) order. We first apply the algorithm to a simple algebraic example and then to the transverse momentum excitations in the holographic theory that describes a stack of M2 branes and includes momentum diffusion as its gapless excitation.
}

\begin{document} 
\maketitle
\flushbottom

\section{Introduction}

Reconstruction of spectra of linear operators \cite{Kato}, given some partial information about its structure, is an old problem. A typical example of such a question in physics may be stated as follows. Imagine a quantum mechanical system with some Hamiltonian that has a corresponding set of energy eigenvalues $E_n$. Now assume that we know one of the eigenvalues as a function of some parameter, such as the coupling constant $\lambda$. For example, we may imagine that we know the ground state energy $E_0(\lambda)$ given as a power series in $\lambda$. Using the knowledge of $E_0$, can one determine the remaining eigenvalues $E_{n\geq 1}(\lambda)$ from $E_0$, thereby reconstructing the full spectrum? Among many works, the problem of this type was considered in a `physically transparent manner' in a series of seminal papers by Bender and Wu \cite{Bender:1968sa,Bender:1969si,Bender:1973rz}. Crucially, those papers explicitly pointed towards the necessity for complexifying the coupling constant $\lambda \in \mathbb{C}$ and showed why complex analytic properties of $E_n$ were essential for answering such questions.\footnote{It is worth noting that, earlier, Dyson also argued for the divergence of perturbative series based on the analytic structure of quantum electrodynamics for `complexified' coupling \cite{PhysRev.85.631}.} In particular, the reconstruction can only follow from the intricate `level-crossing' branch cut structure of the Riemann surface $E_0(\lambda)$ and its relation to the remaining Riemann surfaces $E_{n\geq 1} (\lambda)$.

Since perturbative series in powers of $\lambda$ are typically asymptotic series with zero radii of convergence, these types of problems are intimately related with the field of resurgence in quantum mechanics and quantum field theory (see Refs.~\cite{Dunne:2012ae,Cherman:2013yfa,Dunne:2016jsr,Costin:2020hwg}). Along similar lines, resurgence has also been a fruitful tool for analysing hydrodynamics and its extensions, particularly in the context of holography (see Refs.~\cite{Heller:2011ju,Basar:2015ava,Heller:2015dha,Romatschke:2017vte,Florkowski:2017olj,Strickland:2017kux,PhysRevLett.124.152301,Casalderrey-Solana:2017zyh,Romatschke:2017ejr,Heller:2020jif,Heller:2021oxl,Soloviev:2021lhs}). Among those and related studies, the best understood cases usually pertain to boost invariant (Bjorken-like) flows in `position space' where one complexifies the proper time $\tau\in\mathbb{C}$. What most of the above quantum mechanical, quantum field theoretic, hydrodynamic and holographic studies share is the necessity for analysing asymptotic series, usually by the methods of Pad\'{e} approximants, the Borel resummation and constructions of transseries that contain the knowledge of higher-energy modes.

Instead, one may ask in an analogous manner whether the spectrum of a quantum field theory (QFT) correlation function in momentum (Fourier) space $(\omega, \q)$ can be reconstructed from the knowledge of a single mode's dispersion relation $\omega(q)$, where we define $q\equiv |\q| = \sqrt{\q^2}$. If position space and coupling constant `space' allow for such (at least partial) reconstructions, there is no reason to expect that this should be any different in momentum space. This question was recently considered in Ref.~\cite{Withers:2018srf}, where the reconstruction of the first gapped mode's dispersion relation in the spectrum of a two-point function was successfully worked out from the dispersion relation of a gapless hydrodynamic mode in a holographic model. There, the chosen method of analytic continuation that allowed for the reconstruction was again the Pad\'{e} approximant, however, this time, from the Riemann surface of a gapless $\omega(q)$, which had a convergent series representation around $q=0$.

In this paper, we consider the same type of question. Our first goal is to make a general statement for when a (complete) reconstruction of momentum space dispersion relations contained in correlation functions is possible in theories that exhibit `similar' types of spectra to those seen in holographic models. By `similar' types of spectra, we mean that the correlators only contain poles (are meromorphic) in some complexified parameter space, typically, the frequency. Our second goal is to develop a practical (if somewhat more involved) algorithm that is not based on the method of Pad\'{e} approximants. What we propose is a series of steps that allows for a more direct and rigorous control of all properties of Puiseux series expansions around different critical (branch) points of the Riemann surface, where each step of the reconstruction takes place. Concretely, we will build on a series of recent holographic insights and methods used in Refs.~\cite{Grozdanov:2019kge,Grozdanov:2019uhi,Grozdanov:2021gzh}, which combine two results in complex analysis. Firstly, we will employ the techniques that established, in general, the convergence of classical hydrodynamic dispersion relations through the use of complex spectral curve methods and the Puiseux theorem. Secondly, we will use the theorem of Darboux that can be used to understand all details of a dispersion relation in the vicinity of the critical point limiting its convergence. 

As we will see, for modes in the spectrum that are connected with the same (part of) a non-factorised spectral curve, the reconstruction is possible due to the fact that different physical modes are connected via level-crossings --- i.e., they are all parts of `the same' Riemann surface. On the other hand, when different modes at certain critical points only experience, in the language of \cite{Grozdanov:2019uhi}, `level-touching', then the reconstruction is not possible. This is due to the factorisation of the associated spectral curve.

While our discussion and examples will be closely motivated by holographic theories where our ideas can be explicitly tested, these methods can also be applicable to a variety of theories, most obviously QFTs which allow for a gradient expanded EFT similar to the hydrodynamic expansion. Moreover, we believe that the procedure (the algorithm) that we outline here can be used rather generally in a variety of circumstances when the problem of a spectral reconstruction is considered.

This paper is structured as follows. In Section~\ref{sec:general}, we set up the problem and outline our general strategy and arguments for the reconstruction based on the theorems of Darboux and Puiseux. Then, in Section~\ref{sec:details}, we develop the details of the Darboux theorem and in Section~\ref{sec:examples} consider specific examples that demonstrate the algorithm. Our main physical example is the reconstruction of the spectrum of the transverse momentum, finite temperature retarded two-point function in a holographic large-$N$ 3$d$ conformal field theory (CFT) describing a stack of M2 branes from the structure of the hydrodynamic diffusive mode first studied in Ref.~\cite{Herzog:2002fn}. Finally, in Section~\ref{sec:summary}, we summarise our results and discuss their potential future applications. 

At the end of the paper, we also include three appendices. In Appendix~\ref{app:critical_origin}, we discuss certain specific amendments to the reconstruction algorithm that need to be taken into account when the original dispersion relation is a Puiseux and not a Taylor series. In Appendix~\ref{app2}, we present a variant of the algorithm applicable to cases with two critical points limiting the convergence of the original series. Finally, in Appendix~\ref{app3}, we discuss certain aspects of the reconstruction algorithm using the method of Pad\'{e} approximants instead of the Darboux theorem. There, we also compare the effectiveness of the two methods when applied to the holographic example studied in Section~\ref{sec:M2}.     

\section{The reconstruction: the problem and general strategy}\label{sec:general}

Consider an operator $\CO(t,{\bf x})$ in some $d$-dimensional quantum field theory (QFT) with an associated two-point Green's function in Fourier space. To enable the considerations of thermal states or states with finite density, we allow for the Lorentz invariance to be broken. The Fourier space can then be parametrised with frequency $\omega$ and vector momentum (wavevector) squared $\q^2$, where, for concreteness, we have assumed that the theory remains rotationally invariant under spatial $SO(d-2)$. In what is to follow, it will be essential to consider $\omega$ and $q$, or a frequently used new variable $z \equiv q^2 = \q^2$, as complexified. The Green's function of interest is then $G (\omega, z ) \equiv \langle \CO(\omega, q) \CO(-\omega, -q) \rangle$.

In general, the complex analytic structure of $G(\omega,z)$ in an interacting, potentially strongly coupled QFT can be extremely complicated. The structures in both $\omega$ and $z$ spaces can include infinite numbers of zeros, poles or branch points with a complicated branch cut (Riemann surface) structure (see discussions in Refs.~\cite{Hartnoll:2005ju,Grozdanov:2016vgg,Moore:2018mma,Kurkela:2017xis,Grozdanov:2018gfx}). With a view towards thermal holographic large-$N$ theories, the most obvious prototypical example being the $\CN = 4$ supersymmetric Yang-Mills theory (SYM), here, we will assume that $G(\omega,z)$ is a meromorphic function (without branch cuts) in the complex $\omega$ plane. Note that this assumption is for example violated in a zero temperature CFT. For present purposes, our assumption implies that the spectrum of physical excitations (the modes) is determined by the set of $M$ poles of $G(\omega, z)$, which allows us to write 
\begin{align}
G(\omega, z) = \frac{B(\omega, z)}{\prod_{i=0}^M \left(\omega - \omega_i (z)\right)} .
\end{align}
This set can be, and often is, infinite, i.e.,~$M\to\infty$. On the other hand, $B(\omega, z)$  can have infinitely many zeros but has no additional poles except potentially at infinity.\footnote{Note that for some values of $z$, we can have $B(\omega = \omega_i(z), z) = 0$. These are the pole-skipping points of the correlator $G(\omega,z)$ \cite{Grozdanov:2017ajz,Blake:2017ris,Blake:2018leo,Grozdanov:2018kkt}. For each $i$, there may in fact exist infinitely many such solutions for $z$ \cite{Grozdanov:2019uhi,Blake:2019otz}.} The functions $\omega_i(z)$ are called the {\em dispersion relations} of the modes. 

The problem that we wish to address can now be stated as follows. Assume that we know the dispersion relation of one of the modes, call it $\omega_0(z)$. Can we use the information contained solely in the function $\omega_0(z)$ to find all (or some of the) other dispersion relations $\omega_{i}(z)$, for $i\geq 1$, and thereby reconstruct the entire (or partial) spectrum of $G(\omega,z)$? The fairly obvious answer to this question is: {\it yes}, this is possible for the set of $\omega_i$ that are connected to $\omega_0$ via a sequence of consecutive level-crossings. If this set contains all $\omega_{i}$, for $1\leq i \leq M$ (with $M\to\infty$ possible as well), then we can reconstruct the entire spectrum. 

We now explain more concretely what we mean by the above statement. The relevant language and tools that we will employ can be found in Refs.~\cite{Grozdanov:2019kge,Grozdanov:2019uhi,Grozdanov:2021gzh}. The spectrum of $G(\omega,z)$ follows from its associated complex spectral curve $P(\omega,z)$, with $\omega\in\mathbb{C}$ and $z\in\mathbb{C}$. The zeros of $P(\omega,z)$ are then the poles of the correlator $G(\omega,z)$:\footnote{For example, in the language of holographic calculations, Eq.~\eqref{spectral_curve} is the quasinormal mode condition applied to a gauge invariant mode at the asymptotic anti-de Sitter boundary \cite{Kovtun:2005ev}.} 
\begin{equation}\label{spectral_curve}
    P(\omega=\omega_i(z),z) = 0.
\end{equation}
The spectral curve can also be used to compute the critical points, which satisfy the equation
\begin{equation}\label{critical_point}
    P(\omega=\omega_i(z),z) = 0,~ \partial_\omega P(\omega=\omega_i(z),z) = 0, ~\ldots, ~ \partial^p_\omega P(\omega=\omega_i(z),z) \neq 0. 
\end{equation}
The integer $p$ is called the order of the critical point. In terms of the modes' dispersion relations, critical points are the locations where, out of the full spectrum of modes, two (for $p=2$) or more (for $p > 2$) modes `collide' in the $(\omega,z)$ space. In the language of \cite{Grozdanov:2019uhi}, such collision points can be either of the `level-crossing' or the `level-touching' type. The former, which can be understood as having a non-trivial monodromy, are responsible for the breakdown of convergence of a series representation of $\omega_i(z)$ (for details, see Refs.~\cite{Grozdanov:2019uhi,Grozdanov:2021gzh} and also \cite{Heller:2020uuy}). From the point of view of the analytic structure of dispersion relations, the level-crossing points are the branch points of the Riemann surface and will therefore play a central role in our analysis. 

At a critical point $z_1$ of order $p$, the Puiseux theorem states that there are $p$ solutions to the equation \eqref{spectral_curve}, corresponding to $p$ branches of the Riemann surface defined by \eqref{spectral_curve}, given by convergent Puiseux series
\begin{equation}
\omega_j (z) = \omega(z_1) + \sum_{k\geq k_0}^\infty a_k \left(z-z_1\right)^{k/m_j}, ~~ j = 1,\ldots,p,
\end{equation}
where $m_j$ are positive integers and, in general, $k_0$ can depend on $j$. Crucially, then, if some $m_j > 1$ exists, we will necessarily have among $p$ branches $\omega_j$ a family of $m_j$ solutions of the form
\begin{equation}\label{PuiseuxMjBranches}
\omega_l (z) = \omega(z_1) + \sum_{k\geq k_0}^\infty a_k \left( e^{2\pi i l / m_j} \right)^k \left(z-z_1\right)^{k/m_j}, ~~ l =0,1,\ldots,m_j-1.
\end{equation}
The latter part of the theorem of Puiseux will be of central importance to the entire reconstruction algorithm discussed in this work. 

In practice, having access to any exact dispersion relation is extremely rare. More commonly, we employ the tools of effective field theory (EFT), like hydrodynamics, to write a series representation of $\omega_i(z)$ in powers of $z$. Such an expansion is particularly natural in the low-energy limit since it stems from a derivative (gradient) expansion of relevant fields in position space. By assuming that this is a convenient expansion of the mode $\omega_0(z)$, we can then express its dispersion relation as 
\begin{equation}\label{w0_ser_z0}
    \omega_0(z) = \sum_{n=0}^\infty a_n z^n,
\end{equation}
where the coefficients $a_n$ must be computed from the underlying microscopic theory. What is essential for the usefulness of this procedure is that, for example, in hydrodynamics, the coefficients $a_n$ can be computed as the $\omega \to 0$ and $q\to 0 $ limits of various (fully retarded) higher-point correlation functions by using the analogues of the Kubo formulae and linear response theory. In a thermal QFT, this is incomparably easier than obtaining a two-point function for general $\omega$ and $q$ (see e.g.~Refs.~\cite{Moore:2010bu,Kovtun:2012rj,Grozdanov:2014kva,Grozdanov:2015kqa,Glorioso:2018wxw}). 

In Eq.~\eqref{w0_ser_z0}, we have assumed that the series \eqref{w0_ser_z0} is a Taylor series in $z=q^2$, which amounts to assuming that $(\omega = 0,z=0)$ is a regular point ($p=1$ in \eqref{critical_point}) of the spectral curve. We discuss the relevant techniques and small extensions of our algorithm that are required when the $(\omega = 0,z=0)$ point is a critical point with $p \geq 2$ in Appendix~\ref{app:critical_origin}. Note that from the mathematical point of view, the fact that we assume that $\omega_0$ is given as a series around $z=0$ can be seen as a matter of convenience. All of our results could also be derived from the knowledge of $\omega_0$ expanded around any other point, as well as, clearly, from $\omega_0$ given in its exact form. Note further that in the cases that we consider here, which are motivated by hydrodynamics, $\omega_0$ will be a gapless mode, so $a_0 = 0$. 

The strategy of the reconstruction of $\omega_{i\geq 1}$ then works as follows. The series representation of $\omega_0$ in \eqref{w0_ser_z0} converges in a holomorphic disk of which the radius $R$ is determined by the lowest critical point $z_1$ where level-crossing occurs. That is, $R = |z_1|$. By utilising the theorem of Darboux, which will be discussed in detail in Section~\ref{sec:details}, one can then use the coefficients $a_n$ to compute the location of the critical point $z_1$ (a branch point), its order $p$, and moreover, find the full Puiseux series expansion of $\omega_0(z)$ around $z_1$. Hereon, we will assume that we are dealing with critical points of order $p=2$ and that the Puiseux series exponent corresponding to the reconstructed series is $m_j = p$ (cf.~Eq.~\eqref{PuiseuxMjBranches}). Note that similar statements can also be made for $p > 2$ with $m_j = p$. The Darboux theorem then allows us to find all coefficients $b_n$ of the following Puiseux series:
\begin{equation}\label{w0_ser_z1}
    \omega_0(z) = -i \sum_{n=0}^\infty e^{\frac{i \pi n}{2} } b_n (z-z_1)^{n/2}.
\end{equation}
Since $z_1$ is the point where $\omega_0$ collides with another mode, say $\omega_1$ (i.e.,~$\omega_0(z_1) = \omega_1(z_1)$), one can now use the Puiseux theorem in Eq.~\eqref{PuiseuxMjBranches} to determine the series expansion of the next mode $\omega_1(z)$ around $z_1$ directly from \eqref{w0_ser_z1}. It is given by (for details, see Refs.~\cite{Grozdanov:2019uhi} and \cite{wall})
\begin{equation}\label{w1_ser_z1}
    \omega_1(z) = -i \sum_{n=0}^\infty e^{-\frac{i \pi n}{2} } b_n (z-z_1)^{n/2}.
\end{equation}
The disk of convergence of the series $\omega_1(z)$ in \eqref{w1_ser_z1} is centred at $z_1$ and has the same radius as that of the series \eqref{w0_ser_z1}.

The next steps in the reconstruction of other modes are clear but somewhat difficult to state in a concise manner as they require a `detective'-like approach to exploring the Riemann surfaces of $\omega_0$ and $\omega_1$. In particular, the investigation requires re-expansions and {\it analytic continuations}\footnote{Note that, in principle, any of the numerous known methods of analytic continuation can be chosen.} of $\omega_0$ and $\omega_1$ combined with further uses of the Darboux theorem at critical points to recover $\omega_{i\geq 2}(z)$. Here, we state some options for how this can work. 

What we are required to do is to find a representation of $\omega_0$ or $\omega_1$ that extends at least to the vicinity of the next nearest level-crossing critical point $z_2$ where either $\omega_0$ or $\omega_1$ collides with $\omega_2$. The simplest scenario is that the radius of convergence of \eqref{w1_ser_z1} is limited by $z_2$, in which case one can immediately, as above, use the Darboux theorem to find the Puiseux series representation of $\omega_2(z)$ around $z = z_2$. Another scenario is that a different critical point is limiting the convergence of \eqref{w1_ser_z1}. In the holographic example studied in Section~\ref{sec:M2}, this will be $z^*_1$ (the complex conjugate of $z_1$). One then has two options. The first is to perform analytic continuations of $\omega_0(z)$ in \eqref{w0_ser_z0} or of $\omega_1(z)$ in \eqref{w1_ser_z1} outside their respective radii of convergence to find $z_2$ and reconstruct $\omega_2$ by the Darboux theorem from an appropriate series. Another option, which we develop here, is to use a different ansatz for the Darboux theorem with multiple critical points and directly reconstruct $\omega_2(z)$ around $z_2$ from \eqref{w1_ser_z1}. The success of this step depends on whether $z_2$ is `sufficiently close' to the critical point obstructing \eqref{w1_ser_z1} for the expansions to converge and whether or not there may be even more critical points in the vicinity. We address all those options in Section~\ref{sec:details}. 

Then, with the knowledge of $\omega_0$, $\omega_1$ and $\omega_2$, one proceeds with analogous steps to recursively seek out the remainder of the spectrum. Such successive explorations of the complex $z$ Riemann surfaces of each dispersion relation are expected to eventually recover all dispersion relations $\omega_i(z)$ of the modes that are in any way connected via level-crossings to other, already known, modes. While this general statement is `easy to state', in practice, doing this is a difficult task. In particular, the success depends on the detailed knowledge (e.g.,~the {\it many} coefficients $a_n$) of the behaviour of one of the modes. Moreover, a successful reconstruction also depends on the intricacies of conformal mappings and analytic continuations, which demand a certain level of creativity. Nevertheless, we claim that this procedure is in principle possible, which has important physical implications. In particular, it means that multiple modes --- even all modes in some spectra --- are intimately related and that each one of them has the {\it complete knowledge} of the `physics' of all other modes connected to it via level-crossings. This complete knowledge can extend from the deep infra-red to the extreme ultra-violet energies in the spectrum.  

Finally, it is important to note that, here, we are not claiming a general theorem of complete reconstructability of any QFT correlation function spectrum. This is because the space of functions, theories and examples one can study is infinitely large, and several things could feasibly obstruct the success of such a reconstruction. For example, one could encounter a case when some series would be a lacunary series (an analytic function that cannot be analytically continued), or other issues. Rather, beyond the goal of ours to argue that a complete reconstruction of operator spectra is possible in some examples of highly non-trivial theories of interest (such as in the $\CN = 4$ SYM theory or a lower-dimension theory of M2 branes), another concrete goal is to show how, in practice, this can be constructively done with a controlled method when only a finite number of coefficients $a_n$ is known, for example, from a numerical calculation. We explain the relevant known and new details of the Darboux theorem in the next section and apply them to two examples in Section~\ref{sec:examples}. The main (holographic) example of momentum diffusion in a thermal 3$d$ CFT is studied in Section~\ref{sec:M2}.

\section{The reconstruction: details}\label{sec:details}

In this section, we develop the relevant details of the Darboux theorem that is central to the proposed method of spectral reconstruction in this paper (see e.g.~Ref.~\cite{henrici-book}, Theorem 11.10b). Let us consider a function $f(z)$ that is holomorphic inside a convergent disk $|z| < R$, which we denote by $\mathbb{D}_{R}$. The boundary of the closed disk $\bar{\mathbb{D}}_{R}$ (i.e.,~$|z| \leq R$) will be denoted by $\partial \mathbb{D}_R$. In terms of the modes discussed in Section~\ref{sec:general}, $f(z)$ is the original known dispersion relation in the spectrum, i.e.~$f(z) = \omega_0(z)$. The radius of convergence of the series representation is determined by the critical point of the associated spectral curve $z_1$ where the first level-crossing occurs. This means that $R=|z_1|$. 

The function can in fact have several critical (branch) points located at $|z| = |z_1|$ (i.e.,~at $\partial \mathbb{D}_{|z_1|}$) or in the vicinity of $\partial \mathbb{D}_{|z_1|}$. For this reason, we will structure our discussion into five distinct and relevant cases. First, in Section \ref{1cpt_section}, we will show how the reconstruction algorithm works in the simplest case when there is only a single critical point located at $\partial \mathbb{D}_{|z_1|}$. For successful and fast convergence of the procedure, we will assume that there are no other critical points in the vicinity of $\partial \mathbb{D}_{|z_1|}$. Of course, additional critical points outside $\bar{\mathbb{D}}_{|z_1|}$ are only relevant when it comes to a practical evaluation of the algorithm, not for making formal mathematical statements. Then, in Sections~\ref{sec:2pt_a}, and \ref{sec:2pt_b}, we will consider cases when two critical points $z_1$ and $z_2$ lie precisely at $\partial \mathbb{D}_{|z_1|}$. In Section~\ref{sec:2pt_a}, which is an important case for many physical scenarios, for example, in the holographic calculation in Section~\ref{sec:M2}, we will restrict the two points to being each other's complex conjugates, $z_2 = z_1^*$. In Section~\ref{sec:2pt_c}, we will then use the tools from the cases with two critical points, but apply them to a case with a single critical point at $\partial \mathbb{D}_{|z_1|}$ and another critical point in the vicinity of $\partial \mathbb{D}_{|z_1|}$. This development will significantly improve the practical applicability of our algorithm and this case will also be explicitly used for a part of the calculation in Section~\ref{sec:M2}. Finally, in Section~\ref{sec:2pt_d}, we will consider cases with more than two critical points. The five described scenarios, each requiring a different ansatz in deriving the Darboux theorem, are summarised in Figure~\ref{AO2CP_versions}.

\begin{figure}[h!]%
    \centering
   \hfill
    \subfloat[\centering One critical point at $\partial \mathbb{D}_{|z_1|}$.\label{AO2CP_versions_0}]{{\includegraphics{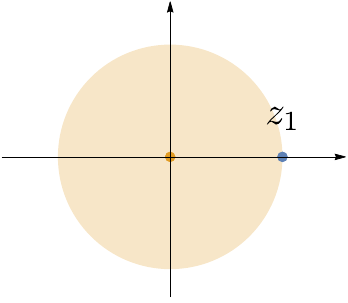} }}
    \hfill
    \subfloat[\centering Two complex conjugated critical points at $\partial \mathbb{D}_{|z_1|}$.\label{AO2CP_versions_a}]{{\includegraphics{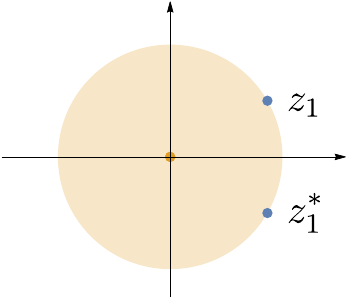} }}
    \hfill
    \label{AO2CP_versions}%
\\
    \centering
    \subfloat[\centering Two general critical points at $\partial \mathbb{D}_{|z_1|}$.\label{AO2CP_versions_b}]{{\includegraphics{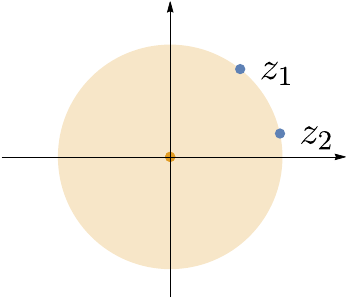} }}
    \hfill
    \subfloat[\centering One critical point at $\partial \mathbb{D}_{|z_1|}$ and another in its vicinity.\label{AO2CP_versions_c}]{{\includegraphics{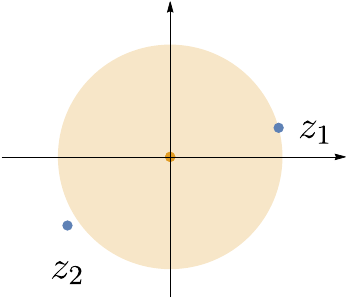} }}
    \hfill
    \subfloat[\centering $J$ critical points at or near $\partial \mathbb{D}_{|z_1|}$.\label{AO2CP_versions_d}]{{\includegraphics{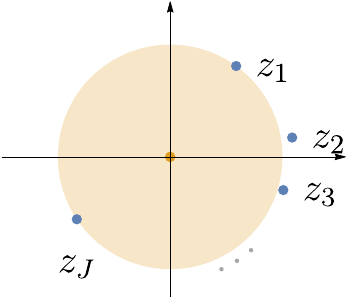} }}
    \hfill
    \caption{Different configurations of critical (branch) points limiting convergence.}%
    \label{AO2CP_versions}%
\end{figure}

In order to develop the reconstruction algorithm in each of the cases, we will assume that we know the coefficients $a_n$ of the convergent power series representation of $f(z)$ in $\mathbb{D}_{|z_1|}$ around the origin:
\begin{equation}\label{f_ser}
f(z) = \sum_{n=0}^\infty a_n z^n.
\end{equation}
The nature of the critical point(s), as well as the expansion of $f(z)$ around $z_1$ can then be worked out from the theorem of Darboux, which relates the asymptotic behaviour of the function $f(z)$ near the critical point $z_1$ to the large-$n$ behaviour of the coefficients $a_n$. However, since, in practice, one rarely knows all coefficients $a_n$, we will also show how the theorem can be implemented given a finite sequence of $a_n$, known analytically or numerically. This procedure will be based on the work of Hunter and Guerrieri \cite{hunter_deducing_1980}.  

Even though the formulae below will be general, as motivated by physical examples, we will mainly think of these critical points as being of order $p=m_j = 2$. In terms of level-crossing, such critical points correspond to `collisions' of a pair of modes in the spectrum \cite{Grozdanov:2019uhi}. Note that, hereon, we will no longer make references to $m_j$ in Eq.~\eqref{PuiseuxMjBranches}. It is, however, important to bear in mind that only the Puiseux series that have the property of $m_j > 1$ can be used in the reconstruction algorithm.

\subsection{One critical point (Case~\ref{AO2CP_versions_0})}
\label{1cpt_section}

We start with the case of $f(z)$ having a single critical point $z = z_1$ on the circle $z = R e^{i\theta}$, where $\theta \in [0,2\pi]$. Without loss of generality, we can take $z_1 \in \mathbb{R}$. In an expansion around $z_1$, the function $f(z)$ has the asymptotic form 
\begin{equation}\label{f_crit_exp}
f(z) \sim \left( z-z_1 \right)^{-\nu} r(z) + \ldots,
\end{equation}
where $r(z)$ is regular in $|z|<\rho$ for some real, positive $\rho>|z_1|$, which therefore allows a convergent expansion with a finite radius of convergence around $z_1$: 
\begin{equation}\label{r_expansion}
r(z)  = \sum_{m=0}^\infty r_m \left(z-z_1\right)^m.
\end{equation}
In terms of the order of the critical point (cf.~Eq.~\eqref{critical_point}),
\begin{equation}
\nu = - 1 / p.
\end{equation}
The ellipsis in \eqref{f_crit_exp} denotes terms that give a subleading contribution to the singularity. For example, for a $p=2$ critical point, which will be of greatest relevance below, we can express all those terms with a single other function $q(z)$, which is also regular at $z=z_1$ and permits an analogous Taylor expansion to $r(z)$: 
\begin{equation}\label{q_expansion}
q(z)  = \sum_{m=0}^\infty q_m \left(z-z_1\right)^m.
\end{equation}
For $p=3$, for example, we would parametrise the subleading (ellipsis) terms with two functions: $(z-z_1)^{2/3} q_1(z)$ and $q_2(z)$. Again, with both $q_1(z)$ and $q_2(z)$ regular at $z = z_1$. 

Let us for the moment keep $\nu$ general. The theorem of Darboux then immediately enables the reconstruction of the leading singular terms in $f(z)$ through the representation of Eq.~\eqref{f_crit_exp}. This is done by using the relation between the asymptotic form of the coefficients $a_n$ for $n\to\infty$ and the coefficients $r_m$. In particular, for $n\to\infty$,
\begin{equation}\label{a_asymp}
a_n \sim \sum_{k=0}^\infty  \frac{(-1)^{k-\nu} (\nu - k)_n r_k}{n! z_1^{n-k+\nu}},
\end{equation} 
where $(k)_n$ denotes the Pochhammer symbol. For a setup with a single critical point at $\partial \mathbb{D}_{|z_1|}$, Eq.~\eqref{a_asymp} is the statement (the result) of Darboux's theorem. The asymptotic form of $a_n$ then allows us to express the exponent $\nu$ in terms of the location the critical point $z_1$:
\begin{equation}\label{nu}
\nu = \lim_{n\to\infty} \left[ z_1 (n+1) \frac{a_{n+1}}{a_n} - n  \right]
\end{equation}
and also gives an iterative expression for all $r_m$ in terms of $r_{0\leq k<m}$ and $a_{n\to\infty}$. Namely,
\begin{equation}\label{r_coeff}
r_m = \lim_{n\to\infty} \left[  \frac{ (-1)^{m-\nu} n! z_1^{n-m+\nu} a_n}{(\nu - m)_n}  - \sum_{k=0}^{m-1} \frac{(-1)^{m-k}(\nu-k)_n r_k}{(\nu-m)_n z_1^{m-k}}  \right].
\end{equation} 
This means that the knowledge of $\nu$ allows to solve  for the critical point location $z_1$ by using the limiting expression in Eq.~\eqref{nu} and $a_{n\to\infty}$. Then, we can iteratively use Eq.~\eqref{r_coeff} to compute {\em all} $r_m$ from the knowledge of $\nu$, $z_1$ and all $a_n$. This constructs the series representation of $r(z)$ in Eq.~\eqref{f_crit_exp}.

It may seem from the above expressions that, in practice, we must know either $z_1$ or $\nu$ in advance in order to be able to proceed and find all $r_m$. However, as we will show below in Section~\ref{sec:one_pt_alg}, this is not necessary. It is sufficient to know only the sequence of coefficients $a_n$ and to avoid the above complication by using the procedure outlined in Ref.~\cite{hunter_deducing_1980}. 

With the complete knowledge of $\nu$, $z_1$ and $r(z)$, we are still left with the task of finding the subleading terms in Eq.~\eqref{f_crit_exp}. For concreteness, let us now fix $\nu = - 1/2$ so that the only remaining function that needs to be found is $q(z)$. The asymptotic form of the original $f(z)$ from Eq.~\eqref{f_crit_exp} is then
\begin{equation}
f(z) \sim \left( z-z_1 \right)^{-\nu} r(z) + q(z)
\label{f_ansatz}
\end{equation}
and the simplest way to find $q(z)$ is by defining a new function $g(z)$ for which $q(z)$ `controls' the square root branch point. One can then use the Darboux theorem for the second time and find the coefficients $q_m$ (cf.~Eq.~\eqref{q_expansion}). Explicitly, we define
\begin{equation}\label{g_fun}
g(z) \equiv \left( z-z_1 \right)^{\nu} f(z) \sim r(z) + \left( z-z_1 \right)^{\nu} q(z),
\end{equation}
so that $g(z)$ has the following a series expansion around $z=0$:
\begin{equation}\label{g_ser}
    g(z)=\sum_{n=0}^\infty g_n z^n ,
\end{equation}
where each coefficient $g_n$ can be found directly from $a_n$:
\begin{equation}
g_n = \sum_{k=0}^n  \frac{(-1)^{n-k-\nu} (\nu)_{n - k} a_k}{(n-k)! z_1^{n-k-\nu}}.
\label{auxiliary_coefficients}
\end{equation}
Clearly, the radius of convergence of the series representation \eqref{g_ser} is the same as that of $f(z)$ in Eq.~\eqref{f_ser}. Next, we can simply use an analogous iterative procedure as for $r_m$. The only difference is that, now, the exponent of $(z-z_1)$ has a different sign. We find
\begin{equation}
q_m=\lim_{n\rightarrow\infty} \left[ \sum_{k=0}^n \frac{(-1)^{n+m-k} n! (\nu)_{n-k} a_k}{(-\nu-m)_n (n-k)! z_1^{m-k}} - \sum_{k=0}^{m-1} \frac{(-1)^{k-m} (-\nu-k)_n q_k}{(-\nu-m)_n z_1^{m-k}} \right] .
\label{AO1CP_even_coeffs_expression}
\end{equation}
In this case, the analogue of the relation \eqref{nu} can be used as a consistency check as one can easily show that
\begin{equation}
\lim_{n\to\infty} \left[ z_1 (n+1) \frac{g_{n+1}}{g_n} - n  \right] = - \lim_{n\to\infty} \left[ z_1 (n+1) \frac{a_{n+1}}{a_n} - n  \right] = - \nu = 1/2. 
\end{equation}

We have thereby reconstructed the full Puiseux series representation of $f(z)$ around the critical point $z = z_1$ of order $p=2$:
\begin{equation}\label{f_ser_CP1}
f(z) =  - i \sum_{n=0}^\infty e^{\frac{i \pi n}{2}} b_n \left(z - z_1\right)^{n/2},
\end{equation}
where in terms of $r_n$ and $q_m$,
\begin{equation}\label{odd_and_even_coefficients}
b_{2n} = i q_n e^{-in\pi} , \qquad b_{2n+1} = r_n e^{-i \left(n+\frac{1}{2}\right)\pi},
\end{equation}
for $n \geq 0$. The expression \eqref{f_ser_CP1} is an analytic continuation of the original series representation of $f(z)$ (cf.~Eq.~\eqref{f_ser}) that was expanded around $z = 0$. The solution in Eq.~\eqref{f_ser_CP1} has the correct structure of a branch of Puiseux series solutions expanded around an order $p=-1/\nu = 2$ critical point. Hence, we can define $f_+ (z) \equiv f(z)$. The second branch then follows from the general structure of Puiseux series (see Refs.~\cite{Grozdanov:2019uhi} and \cite{wall}). In particular, the two solutions written together are (cf.~Eq.~\eqref{w1_ser_z1}, which follows from Eq.~\eqref{PuiseuxMjBranches}) 
\begin{equation}\label{f_ser_both}
f_\pm (z) =  - i \sum_{n=0}^\infty e^{\pm\frac{ i \pi n}{2}} b_n \left(z - z_1\right)^{n/2}.
\end{equation}
In terms of the modes discussed in Section~\ref{sec:general}, $f_-(z) = \omega_1(z)$ is the second solution. 

Finally, we note that a similar procedure of redefining the original function $f(z)$ can also be performed in order to isolate different regular functions for $p \geq 3$. For example, for $p=3$, these were called $q_1(z)$ and $q_2(z)$.

\subsubsection{The algorithm for calculating the series}\label{sec:one_pt_alg}

Next, we ask the question of how the above procedure can be executed in practice when only a finite number of the series coefficients $a_n$ is known; e.g.,~for some $0\leq n \leq N$. As mentioned above, this can be done following the procedure proposed by Hunter and Guerrieri in Ref.~\cite{hunter_deducing_1980}. In particular, we define the following recurrence relation constructed purely in terms of the known coefficients $a_n$:
\begin{equation}
\begin{aligned}
X^0_n (\nu, z_1) &= a_n , \\
X^{m+1}_n (\nu, z_1) &= X^m_n (\nu, z_1) - \frac{(n+\nu-2m-1)}{n z_1} X^m_{n-1} (\nu, z_1) , ~~ \text{for}~m \geq 0.
\end{aligned}
\label{AO1CP_X_definition}
\end{equation}
Crucially, the asymptotic expansion of $X^m_n$ is given by 
\begin{equation}\label{asymp_HG_1}
X^m_n (\nu, z_1)  \sim \sum_{k=m}^\infty \frac{(-1)^{k+m-\nu} k! (\nu-k)_{n-m} r_k}{n! (k-m)! z_1^{n+\nu-k}} \sim O(n^{\nu-2m-1} ).
\end{equation}
Hence, in the limit of large $n$ and large $m$, the right-hand-side of \eqref{asymp_HG_1} tends to zero. Choosing two different $X^m_n$ and setting them to zero (up to errors, which tend to zero as $n,m\to\infty$), we can then solve a simultaneous system of two equations for two unknowns: $\nu$ and $z_1$. It is most practical to first set 
\begin{equation}
X^1_{N} = 0, \qquad X^1_{N - 1} = 0,
\end{equation}
so that the system of equations is linear and $\nu$ and $z_1$ unique. Then, we can use these solutions as seeds to numerically iteratively solve for 
\begin{equation}
X^m_{N} = 0, \qquad X^m_{N - 1} = 0,
\label{bp_algo_HG_1}
\end{equation}
with increasing $m\leq M$, for some $M$. At each step, we look for solutions close to $\nu$ and $z_1$. Due to the rapid convergence of this procedure, we can easily check whether the iteratively generated solutions are correct. Thereby, as noted above, we indeed recover both $\nu$ and $z_1$ purely from the knowledge of $a_n$. 

Next, we extend the discussion of \cite{hunter_deducing_1980} to also find the coefficients $r_m$ of the series \eqref{r_expansion}. In particular, we define a different recurrence relation 
\begin{equation}
\begin{aligned}
Y^0_{\ell,n} (\nu, z_1) &= a_n , \\
Y^{m+1}_{\ell,n} (\nu, z_1) &= Y^m_{\ell,n} (\nu, z_1) - \frac{(n+\nu-2m-\ell - 2)}{n z_1} Y^m_{\ell, n-1} (\nu, z_1) , ~~ \text{for}~m \geq 0.
\label{AO1CP_Y_definition}
\end{aligned}
\end{equation}
It is easy to check that this recurrence naturally gives $r_0$ when we choose $\ell = 0$. Then, iteratively, recurrences with higher $\ell$ lead to the coefficients $r_\ell$. To see this, first take $\ell = 0$. It can then be shown that the leading term in the asymptotic expansion of $Y^m_{\ell = 0 ,n}$ is
\begin{equation}\label{asymp_HG_2}
Y^m_{0,n} (\nu, z_1)  \sim  \frac{(-1)^{-\nu} m! (\nu)_{n-m} r_0}{n! z_1^{n+\nu}} + O(n^{\nu-2m-2}).
\end{equation}
Hence, in the limit of large $n$ and large $m$, the expression \eqref{asymp_HG_2} indeed allows us to accurately compute $r_0$ from the known $\nu$ and $z_1$. For general $\ell$, 
\begin{equation}
Y_{\ell,n}^m \sim \sum_{k=0}^\ell \frac{(-1)^{k-\nu} (m+\ell-k)! (\nu-k)_{n-m} r_k}{n! (\ell-k)!z_1^{n+\nu-k}} + \mathcal{O}(n^{\nu - 2m-\ell-2}),
\label{AO1CP_Y_asymptotic_form}
\end{equation}
from where the expression for $r_\ell$ given in terms of the coefficients $r_{0\leq k < \ell}$ follows:
\begin{equation}
r_\ell = \lim_{n\rightarrow\infty} \left[ \frac{(-1)^{\ell-\nu} n! z_1^{n+\nu-\ell}}{ m! (\nu-\ell)_{n-m}}Y_{\ell,n}^m - \sum_{k=0}^{\ell-1} \binom{m+\ell-k}{m} \frac{(-1)^{\ell-k}(\nu-k)_{n-m}r_k}{(\nu-\ell)_{n-m}z_1^{\ell-k}} \right].
\label{coeffs_algo_HG_1}
\end{equation}

To obtain the subleading (even) coefficients $q_k$ of the series \eqref{q_expansion}, one employs the same algorithm using the coefficients $g_n$ defined in \eqref{auxiliary_coefficients}.

\subsubsection{A very simple example: a quadratic complex algebraic curve}\label{sec:ex_analytical}

Before continuing with the exposition of more complicated cases of Darboux's theorem, we show how the above procedure for cases with a single critical point can be implemented on an example of a simple quadratic spectral curve. Due to its simplicity, we will be able to analyse this case analytically. The complex spectral curve that we consider here is 
\begin{equation}
P(\omega,z) = \omega^2+\omega-z=0.
\end{equation}
The equation has two solutions representing the two branches of the Riemann surface:
\begin{equation}
\omega_{\pm} (z)= -\frac{1}{2} \pm \sqrt{z-z_1},
\label{SimpleAnalyticExampleSolution}
\end{equation}
with $z_1=-1/4$ being the only critical point. Note that this can be shown by using Eq.~\eqref{critical_point}. Of the two solutions, in the language of physical modes, $\omega_+(z)$ is the {\it gapless} `hydrodynamic' mode: $\omega_+(0)=0$. The second mode is {\it gapped}: $\omega_-(0)=-1$.

To demonstrate the proposed reconstruction procedure, we imagine that we are only given the coefficients $a_n$ of the Taylor series expansion of $f(z) = \omega_+(z)$ around the point $z=0$:
\begin{equation}\label{f_quad}
f(z)= \sum_{n=1}^\infty a_n z^n.
\end{equation}
Explicitly, the coefficients are
\begin{equation}
a_n = 2^{2n-1}\binom{1/2}{n}.
\end{equation}
One can easily check that the radius of convergence of the series \eqref{f_quad} is indeed $1/4$. 

Our first task is now to find the location of the critical point $z_1$ and the exponent $\nu$ of the asymptotic form of $f(z) \sim (z-z_1)^{-\nu} r(z)$. Since $\nu$ will turn out to be equal to $-1/2$, we will then be able to compute the Puiseux series coefficients $b_n$ of the expansion 
\begin{equation}
f(z) \equiv f_+(z) =  - i \sum_{n=0}^\infty e^{\frac{i \pi n}{2}} b_n \left(z - z_1\right)^{n/2}
\end{equation}
around $z_1$, and, finally, use them to determine the series expansion of the second branch, which will correspond to $\omega_-(z)$ (cf.~Eq.~\eqref{f_ser_both}):
\begin{equation}
f_- (z) =  - i \sum_{n=0}^\infty e^{-\frac{ i \pi n}{2}} b_n \left(z - z_1\right)^{n/2}.
\end{equation}

We start by constructing the polynomials $X^m_n$, which are defined in Eq.~\eqref{AO1CP_X_definition}. For $m=1$, we get
\begin{equation}
X^1_n = -\frac{2^{2n-3}}{n z_1} \left[\left(4z_1+1\right)n-6z_1+\nu-1\right] \binom{1/2}{n-1}.
\end{equation}
Solving the system of equations $X^1_n=0$ and $X^1_{n-1}=0$ for $\nu$ and $z_1$, one immediately obtains the correct result:
\begin{equation}
z_1=-\frac{1}{4}  \quad\text{and}\quad \nu=-\frac{1}{2}.
\end{equation}
In this case, there is no need to use higher $m$. The result is exact and analytical. In more complicated examples considerer below, however, we will see that the accuracy of numerical calculations benefits enormously from a recursive calculation at higher and higher $m$.

Next, we determine the coefficients $b_n$, which is done in two steps. First, we compute the odd-$n$ and then, the even-$n$ coefficients (cf.~Eq.~\eqref{odd_and_even_coefficients}). By using Eq.~\eqref{AO1CP_Y_definition} to construct the polynomials $Y^m_{\ell,n}$ and setting $m=1$ and $\ell=0$, we find
\begin{equation}
Y^1_{0,n} = - \frac{4^{n-1}\sqrt{\pi}}{\Gamma(n+1)\Gamma(5/2-n)}.
\end{equation}
Using Eq.~\eqref{coeffs_algo_HG_1} for the coefficients $r_\ell$, it follows that
\begin{equation}
r_0 = \lim_{n\rightarrow\infty}\left[ \frac{(-1)^{-\nu} n! z_1^{n+\nu}}{m! (\nu)_{n-m}} Y^1_{0,n} \right] = 1,
\end{equation}
while all other $r_\ell$ vanish. This fixes all coefficients $b_n$ with odd $n$. 

Next, we determine the coefficients $b_n$ with even $n$ by utilising the auxiliary function $g(z)$, as defined in Eq.~\eqref{g_fun}. The coefficients of its series expansion around $z=0$ are given by \eqref{auxiliary_coefficients}. In this case, we find that
\begin{equation}
g_n = -4^n \binom{-\frac{1}{2}}{n}.
\end{equation}
One can check that the branch point of $g(z)$ is at the same location as the branch point of the original function $\omega_+(z)$, i.e., at $z = z_1=- 1/4$, while the relevant `asymptotic scaling' exponent is now $-\nu = 1/2$ instead of $\nu = -1/2$. We then construct the polynomials $Y^1_{0,n}$,
\begin{equation}
Y^1_{0,n} = \frac{4^{n}\sqrt{\pi}}{\Gamma(n+1)\Gamma(3/2-n)},
\end{equation}
and find that
\begin{equation}
q_0 = \lim_{n\rightarrow\infty}\left[ \frac{(-1)^{+\nu} n! z_1^{n-\nu}}{m! (-\nu)_{n-m}} Y^1_{0,n} \right] = -\frac{1}{2},
\end{equation}
while all other $q_\ell$ vanish. This fixes all coefficients $b_n$ with even $n$. 

By using Eqs.~\eqref{r_expansion} and \eqref{q_expansion} (along with Eq.~\eqref{odd_and_even_coefficients}) in \eqref{f_ansatz}, we obtain the correct series expansion of $f = \omega_+$ around $z_1$:  
\begin{align}
f(z) &\equiv f_+(z) = \omega_+(z) = (z-z_1)^{-\nu} \sum_{m=0}^\infty r_m \left(z-z_1\right)^m + \sum_{m=0}^\infty q_m \left(z-z_1\right)^m \nn
& =  - \frac{1}{2} + \sqrt{z+\frac{1}{4}}. \label{SP1}
\end{align}
Finally, this result immediately gives us the Puiseux series expansion of the second solution $f_-(z) = \omega_-(z)$ around $z = z_1$ (cf.~Eq.~\eqref{f_ser_both}):
\begin{equation}\label{SP2}
f_-(z) = \omega_-(z) = - \frac{1}{2} -\sqrt{z+\frac{1}{4}}.
\end{equation}
Note that the Puiseux series expansions of $f_\pm$ around the critical point $z_1$ are finite series. Hence, they both have an infinite radius of convergence unlike the Taylor series representation of $f_+$ around $z=0$. The reason is that the Puiseux series representations in Eqs.~\eqref{SP1} and \eqref{SP2} happen to be the exact solutions \eqref{SimpleAnalyticExampleSolution} of the spectral curve equation. The two solutions represent the entire Riemann surface associated with the spectral curve. 

\subsection{Two complex conjugated critical points (Case~\ref{AO2CP_versions_a})}
\label{sec:2pt_a}

Next, we consider cases with two critical (branch) points located at the boundary of the convergence disk $\bar{\mathbb{D}}_{|z_1|}$ (i.e.,~$\partial \mathbb{D}_{|z_1|}$) that limit the convergence of \eqref{f_ser}. In particular, in this subsection, we will start with scenarios in which the two critical points $z_1$ and $z_2$ are each other's complex conjugates: $z_2 = z_1^*$. We depict this situation in Figure~\ref{AO2CP_versions_a}. We note that in Ref.~\cite{Grozdanov:2021gzh}, such a case was considered by combining the Darboux theorem with a conformal map (a M\"{o}bius transform), which moved one of the critical points away from the boundary of the convergence disk. If such a conformal map can be found in practice (which is highly non-trivial and potentially impractical as it can drastically impair the convergence of the algorithm), then one can proceed by using the `one critical point' algorithm described in Section~\ref{1cpt_section}. 

In all cases with two critical points, in addition to the fact the function $f(z)$ has the asymptotic form
\begin{equation}
f(z) \sim \left(z-z_1\right)^{-\nu} r(z) + q(z)
\label{f_asymp_2cp}
\end{equation}
as $z\rightarrow z_1$, we also have that
\begin{equation}
f(z) \sim \left(z-z_2\right)^{-\nu} p(z) + s(z)
\end{equation}
as $z\rightarrow z_2$. Here, $r(z)$, $q(z)$, $p(z)$ and $s(z)$ are again all regular in $|z|<\rho$ for some $\rho>|z_1|$.

Asymptotic behaviour of the coefficients $a_n$ of the expansion of $f(z)$ around the origin is now given by the sum of two series of the form \eqref{a_asymp} for the two critical points:
\begin{equation}
a_n \sim \sum_{k=0}^\infty  \frac{(-1)^{k-\nu} (\nu - k)_n}{n! R^{n-k+\nu}} \left( r_k e^{-i(n-k+\nu)\theta} + p_k e^{i(n-k+\nu)\theta} \right) ,
\label{a_asymp_2cp}
\end{equation}
where we designate the coefficients of the expansion of $r(z)$ and $p(z)$ around $z_1$ and $z_2$, respectively, as $r_k$ and $p_k$, while we write the two critical points as $z_1=z_2^*=Re^{i\theta}$.

Here, we also note that another way to perform this analysis is through a more `symmetric' representation of the function $f(z)$ given by
\begin{equation}
f(z) = \left(z-z_1\right)^{-\nu} \left(z-z_2\right)^{-\nu} R(z) + Q(z),
\label{a_asymp_2cp_Gegenbauer}
\end{equation}
where we can then make use of the Taylor multi-point expansion of $R(z)$:
\begin{equation}
R(z) = \sum_{k=0}^\infty (\alpha_k + z \beta_k) (z-z_1)^k (z-z_2)^k.
\end{equation}
In some cases, this (formally equivalent) ansatz may be more useful than the one discussed above, but since this had not proved to be so in our numerical calculations, we delegate the details of its presentation to Appendix~\ref{app2}.

\subsubsection{The algorithm for calculating the series}

We continue by using the form \eqref{a_asymp_2cp} and extend the work done in \cite{hunter_deducing_1980} by defining the following recursion relation in terms of three unknowns $\nu$, $R$ and $\cos\theta$:
\begin{equation}
\begin{aligned}
X_n^0(\nu, R, \cos\theta) &= a_n , \\
X_n^{m+1}(\nu, R, \cos\theta) &= X_n^{m}(\nu, R, \cos\theta) - 2\cos\theta \frac{n+\nu-2m-1}{n R} X_{n-1}^m(\nu, R, \cos\theta) \\
&\phantom{ = }+ \frac{(n+\nu-m-1)(n+\nu-3m-2)}{n(n-1)R^2} X_{n-2}^m(\nu, R, \cos\theta),
\end{aligned}
\end{equation}
which asymptotically scale as
\begin{equation}
X_n^m \sim \mathcal{O}(n^{\nu-3m-1}).
\end{equation}
This means that we can solve for $\nu$, $R$ and $\theta$ by setting to zero three consecutive polynomials at $m=1$:
\begin{equation}
X_N^1 = 0, \quad X_{N-1}^1=0, \quad X_{N-2}^1=0.
\end{equation}
We then iterate the procedure in increasing $m\leq M$, for some $M$, at each step solving the three equations
\begin{equation}\label{ao2cp_bp_algorithm}
X_N^m = 0, \quad X_{N-1}^m=0, \quad X_{N-2}^m=0,
\end{equation}
and taking the previous solution in $m$ as the seed for the next one.

Next, we calculate the coefficients (again extending \cite{hunter_deducing_1980}) $r_\ell$ and $p_\ell$ in \eqref{a_asymp_2cp} by recursively defining the polynomials $Y_{\ell,n}^m$ as
\begin{equation}
\begin{aligned}
Y_{\ell,n}^0 &= a_n ,\\
Y_{\ell,n}^{m+1} &= Y_{\ell,n}^{m} - 2\cos\theta \frac{n+\nu-2m-\ell-2}{n R} Y_{\ell,n-1}^m \\
&\phantom{ = }+ \frac{(n+\nu-m-\ell-2)(n+\nu-3m-\ell-3)}{n(n-1)R^2} Y_{\ell,n-2}^m ,
\end{aligned}
\label{AO2CP_Y_recursion}
\end{equation}
which behave as
\begin{equation}
Y_{\ell,n}^m = \sum_{k=0}^\ell \mathcal{F}^m_{n,k} + \mathcal{O}(n^{\nu-3m-\ell-2}),
\end{equation}
where $\mathcal{F}^m_{n,k}$ is an expression linear in $r_k$ and $p_k$ that follows directly from \eqref{AO2CP_Y_recursion}. In practice, $\mathcal{F}^m_{n,k}$ is easy to compute and we evaluate it in the process of the calculation. For conciseness, we will not state its explicit form at any of the steps here. 

To obtain the $\ell$-th coefficients $r_\ell$ and $p_\ell$ in terms of $r_{0\leq k < \ell}$ and $p_{0\leq k< \ell}$, we then consider the linear system of equations
\begin{equation}\label{ao2cp_coeffs_algorithm}
Y_{\ell,N}^M - \sum_{k=0}^\ell \mathcal{F}^M_{N,k}=0, \quad Y_{\ell,N-1}^M - \sum_{k=0}^\ell \mathcal{F}^M_{N-1,k}=0,
\end{equation}
where the error scales as $\mathcal{O}(N^{\nu-3M-\ell-2})$ at large $N$, and solve the system of equations for the two unknowns $r_\ell$ and $p_\ell$.

Again, in order to determine the coefficients of the function $q(z)$ from \eqref{f_asymp_2cp}, we, as in the case of a single critical point, define an auxiliary function $g(z)$ by
\begin{equation}
g(z) \equiv \left( z-z_1 \right)^\nu f(z) \sim r(z) + (z-z_1)^\nu q(z),
\end{equation}
and use the above algorithm \eqref{ao2cp_coeffs_algorithm} on the coefficients $g_n$, which are again given by
\begin{equation}
g_n = \sum_{k=0}^n  \frac{(-1)^{n-k-\nu} (\nu)_{n - k} a_k}{(n-k)! z_1^{n-k-\nu}}.
\end{equation}

In this manner, we reconstruct the full Puiseux series representation of $f(z)$ around $z_1$, one of the two closest critical points,
\begin{equation}
f(z) =  - i \sum_{n=0}^\infty e^{\frac{i \pi n}{2}} b_n \left(z - z_1\right)^{n/2},
\end{equation}
where $b_n$ can be written in terms of $r_n$ and $q_n$ as $b_{2n} = i q_n e^{-in\pi}$ and $b_{2n+1} = r_n e^{-i\left(n+\frac{1}{2}\right)\pi}$. Finally, the expansions of the two branches around the critical point are the two Puiseux series given in Eq.~\eqref{f_ser_both}.

\subsection{Two general critical points (Case~\ref{AO2CP_versions_b})}\label{sec:2pt_b}

If the obstruction to the convergence of a given series is caused by two critical points $z_1$ and $z_2$ that are equal distance away from the centre of the expansion, but are not conjugate to each other (as in Figure~\ref{AO2CP_versions_b}), then we use the following ansatz:
\begin{equation}
a_n \sim \sum_{k=0}^\infty  \frac{(-1)^{k-\nu} (\nu - k)_n}{n! R^{n-k+\nu}} \left( r_k e^{-i(n-k+\nu)\theta_1} + p_k e^{i(n-k+\nu)\theta_2} \right) ,
\end{equation}
where $z_1=R e^{i \theta_1}$ and $z_2=R e^{i \theta_2}$.

We again use the two recursively defined polynomials $X_n^m$ and $Y_{\ell,n}^m$ for the calculation of the branch point position and the coefficients of the expansion, respectively. They are defined by the following expressions:
\begin{equation}
\begin{aligned}
X_n^0(\nu, R, \theta_1, \theta_2) &= a_n ,\\
X_n^{m+1}(\nu, R, \theta_1, \theta_2) &= R^2 e^{i(\theta_1 + \theta_2)} X_n^{m}(\nu, R, \theta_1, \theta_2) \\
&\phantom{ = }- R \left( e^{i\theta_1} + e^{i\theta_2} \right) \frac{n+\nu-2m-1}{n} X_{n-1}^m(\nu, R, \theta_1, \theta_2) \\
&\phantom{ = }+ \frac{(n+\nu-m-1)(n+\nu-3m-2)}{n(n-1)} X_{n-2}^m(\nu, R, \theta_1, \theta_2),
\end{aligned}
\end{equation}
and
\begin{equation}
    \begin{aligned}
    Y_{\ell,n}^0 &= a_n ,\\
    Y_{\ell,n}^{m+1} &= R^2 e^{i(\theta_1 + \theta_2)} Y_{\ell,n}^{m} - R \left( e^{i\theta_1} + e^{i\theta_2} \right) \frac{n+\nu-2m-\ell-2}{n} Y_{\ell,n-1}^m \\
    &\phantom{ = }+ \frac{(n+\nu-m-\ell-2)(n+\nu-3m-\ell-3)}{n(n-1)} Y_{\ell,n-2}^m.
    \end{aligned}
\end{equation}
Setting, for $m=1,\dots,M$,
\begin{equation}
    X_N^m = X_{N-1}^m = X_{N-2}^m = X_{N-3}^m = 0,
\end{equation}
then allows us to calculate $\nu$, $R$, $\theta_1$ and $\theta_2$, while from
\begin{equation}
    Y_{\ell,N}^M - \sum_{k=0}^\ell \mathcal{F}^M_{N,k} = 0, \quad Y_{\ell,N-1}^M - \sum_{k=0}^\ell \mathcal{F}^M_{N-1,k} = 0, 
\end{equation}
we calculate the coefficients $r_\ell$ and $p_\ell$, provided the $r_{0\leq k < \ell}$ and $p_{0\leq k < \ell}$ are  known at each iterative step. Again, the expressions $\mathcal{F}^m_{n,k}$ are calculated in the process of the evaluation of the recursive algorithm.

\subsection{One closest critical point with another in its vicinity (Case~\ref{AO2CP_versions_c})}
\label{sec:2pt_c}

In practice, even if only one critical point $z_1$ formally obstructs the convergence of the series, but the next closest critical point $z_2$ lies in the vicinity of the boundary of the convergence disk $\partial \mathbb{D}_{|z_1|}$, we find that it is vastly beneficial to use the ansatz with two critical points from Section~\ref{sec:2pt_b} (Case~\ref{AO2CP_versions_b}):
\begin{equation}
a_n \sim \sum_{k=0}^\infty  \frac{(-1)^{k-\nu} (\nu - k)_n}{n!} \left( \frac{r_k}{z_1^{n-k+\nu}} + \frac{p_k}{z_2^{n-k+\nu}} \right),
\end{equation}
where we have kept $z_1$ and $z_2$ general. In this way, the convergence of the Darboux theorem is greatly improved compared to the algorithm that only takes into account the single closest critical point $z_1$. Now, the appropriate expressions for $X_n^m$ and $Y_{\ell,n}^m$ are defined as
\begin{equation}
\begin{aligned}
X_n^0(\nu, z_1, z_2) &= a_n, \\
X_n^{m+1}(\nu, z_1, z_2) &= z_1 z_2 X_n^{m}(\nu, z_1,z_2) - \frac{n+\nu-2m-1}{n} \left(z_1 + z_2 \right) X_{n-1}^m(\nu, z_1, z_2) \\
&\phantom{ = }+ \frac{(n+\nu-m-1)(n+\nu-3m-2)}{n(n-1)} X_{n-2}^m(\nu, z_1, z_2),
\end{aligned}
\end{equation}
and
\begin{equation}
    \begin{aligned}
    Y_{\ell,n}^0 &= a_n, \\
    Y_{\ell,n}^{m+1} &= z_1 z_2 Y_{\ell,n}^{m} - \left( z_1 + z_2 \right) \frac{n+\nu-2m-\ell-2}{n} Y_{\ell,n-1}^m \\
    &\phantom{ = }+ \frac{(n+\nu-m-\ell-2)(n+\nu-3m-\ell-3)}{n(n-1)} Y_{\ell,n-2}^m. 
    \end{aligned}
\end{equation}
Setting, for $m=1,\dots,M$,
\begin{equation}
    X_N^m = X_{N-1}^m = X_{N-2}^m = X_{N-3}^m = X_{N-4}^m = 0 ,
    \label{AO2CP_NC_NE}
\end{equation}
allows us to calculate $\nu$, $z_1$ and $z_2$, while from
\begin{equation}
    Y_{\ell,N}^M - \sum_{k=0}^\ell \mathcal{F}^M_{N,k} = 0, \quad Y_{\ell,N-1}^M - \sum_{k=0}^\ell \mathcal{F}^M_{N-1,k} = 0,
    \label{AO2CP_NC_NE_coeffs}
\end{equation}
we calculate the coefficients $r_\ell$ and $p_\ell$, provided that the $r_{0\leq k < \ell}$ and $p_{0\leq k < \ell}$ are known. The rest of the procedure continues as before.

\subsection{$J$ critical points with $J\geq 3$ (Case~\ref{AO2CP_versions_d})}
\label{sec:2pt_d}

In general, one can use an ansatz with as many branch points as one believes may be useful in a given situation. Even though, in practice, this is a difficult thing to assess given only the knowledge of $a_n$, we nevertheless state the relevant results of such a procedure here. The ansatz for $J$ points reads as
\begin{equation}
a_n \sim \sum_{k=0}^\infty  \frac{(-1)^{k-\nu} (\nu - k)_n}{n!} \sum_{j=1}^J \frac{r^{(j)}_k}{z_j^{n-k+\nu}},
\end{equation}
with the $X_n^m$ and $Y_{\ell,n}^m$ defined by
\begin{equation}
\begin{aligned}
X_n^0(\nu, S) &= a_n , \\
X_n^{m+1}(\nu, S) &= \sum_{i=0}^J (-1)^i \left( \sum_{A\in S_k} \prod_{z\in A} z \right) \alpha_i(\nu,n,m) X_{n-i}^m(\nu, S),
\end{aligned}
\end{equation}
and
\begin{equation}
\begin{aligned}
Y_{\ell,n}^0(\nu, S) &= a_n ,\\
Y_n^{m+1}(\nu, S) &= \sum_{i=0}^J (-1)^i \left( \sum_{A\in S_k} \prod_{z\in A} z \right) \beta_i(\nu,n,m) X_{n-i}^m(\nu, S).
\end{aligned}
\label{JCP_bp_recursion}
\end{equation}
Here,
\begin{equation}
\begin{aligned}
    \alpha_i(\nu,n,m)&=\frac{(1+(1-\delta_{i,0})(n+\nu-(i+1)m-i-1))\prod_{j=2}^i (n+\nu-m-j+1)}{\prod_{j=1}^i (n-j+1)}, \\
    \beta_i(\nu,n,m)&=\frac{(1+(1-\delta_{i,0})(n+\nu-(i+1)m-i-1))\prod_{j=2}^i (n+\nu-m-j+1)}{\prod_{j=1}^i (n-j+1)} ,
\end{aligned}
\end{equation}
and we used $S=\{z_j\}_{j=1}^J$ to denote the set of all ($J$) included critical points. $S_k$ is then the set of all possible $k$-combinations of the elements of $S$.

Taking, for example, $J=4$, the recursive definition of $X_n^m$ in \eqref{JCP_bp_recursion} becomes
\begin{equation}
    \begin{aligned}
        X^{m+1}_n &= z_1z_2z_3z_4 X_n^m - \left( z_1z_2z_3 + z_1z_2z_4 + z_1z_3z_4 + z_2z_3z_4 \right) \frac{n+\nu-2m-1}{n} X_{n-1}^m \\
        &+ \left(z_1z_2 + z_1z_3 + z_1z_4 + z_2z_3 + z_2z_4 + z_3z_4\right)\frac{(n+\nu-m-1)(n+\nu-3m-2)}{n(n-1)}X_{n-2}^m \\
        &- (z_1+z_2+z_3+z_4)\frac{(n+\nu-m-1)(n+\nu-m-2)(n+\nu-4m-3)}{n(n-1)(n-2)}X_{n-3}^m \\
        &+ \frac{(n+\nu-m-1)(n+\nu-m-2)(n+\nu-m-3)(n+\nu-5m-4)}{n(n-1)(n-2)(n-3)} X_{n-4}^m ,
    \end{aligned}
\end{equation}
and one can similarly construct the expression for $Y_{\ell,n}^m$.

\section{The reconstruction: examples}\label{sec:examples}
\subsection{A simple cubic complex algebraic curve}
\label{ao1cp_example}

\begin{figure}[h!]
    \centering
    \includegraphics[scale=0.95]{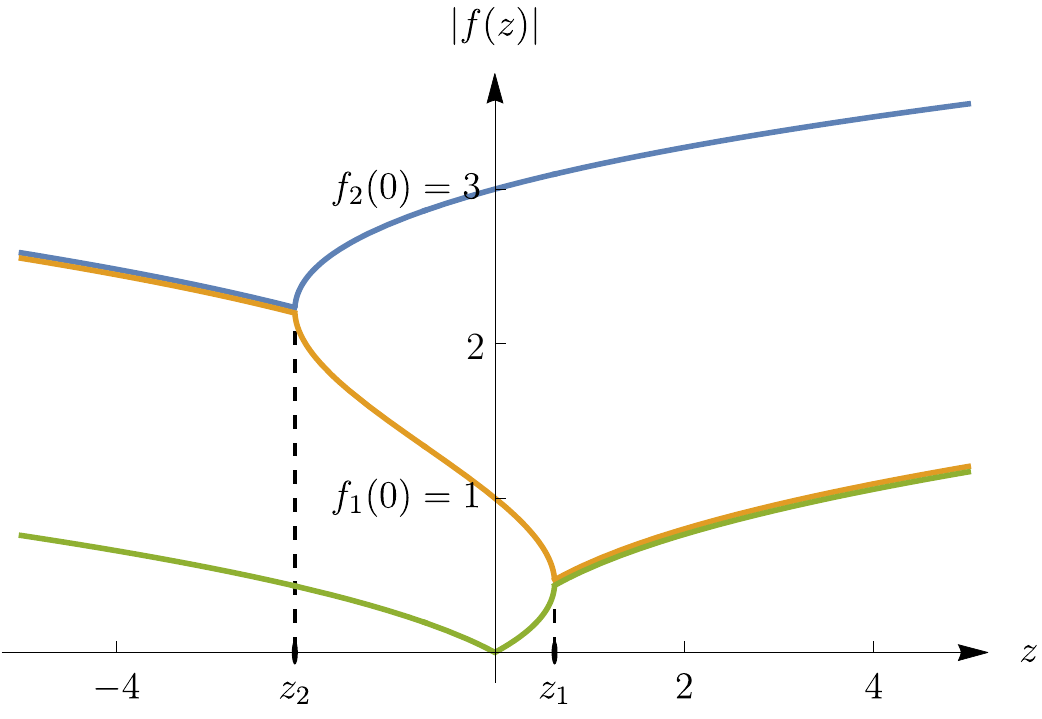}
    \caption{Absolute values of the solutions $f_0(z)$ (in green), $f_1(z)$ (in orange) and $f_2(z)$ (in blue) to Eq.~\eqref{P_alg_ex} plotted for real $z$. The two real critical points are labeled by $z_1$ and $z_2$ (cf.~Eq.~\eqref{P_alg_ex_cp}).}
    \label{fig:AO1CP_made_up_example}
\end{figure}

To employ our general approach to the reconstruction of spectra and demonstrate the utility of the above algorithms, beyond the very simple example from Section~\ref{sec:ex_analytical}, which could be treated analytically, we first consider the following simple cubic algebraic spectral curve:
\begin{equation}\label{P_alg_ex}
P(f,z) = 3 f - 4 f^2 + f^3 - z = 0,
\end{equation}
where we treat $f$ and $z$ as complex variables. The three solutions of Eq.~\eqref{P_alg_ex}, $f_{0}(z)$, $f_1(z)$ and $f_2(z)$, are easy to obtain in closed form and will serve as a check of various steps in the course of the reconstruction. The spectrum has two (real) critical points of order $p=2$ that satisfy Eq.~\eqref{critical_point}:
\begin{equation}\label{P_alg_ex_cp}
\begin{aligned}
    z_1 &= \frac{2}{27} \left(7\sqrt{7} - 10 \right), & f(z_1) &= \frac{1}{3} \left(4- \sqrt{7}\right) ,\\ 
    z_2 &= -\frac{2}{27} \left( 7\sqrt{7} + 10 \right), & f(z_2) &= \frac{1}{3} \left(4 +  \sqrt{7}\right).
\end{aligned}
\end{equation}
We depict the three-sheeted Riemann surface structure of solutions by plotting $|f_0(z)|$, $|f_1(z)|$ and $|f_2(z)|$ for real $z$ in Figure~\ref{fig:AO1CP_made_up_example}.

To show how the reconstruction of the full spectrum from a single mode $f_0$ works, we assume we know a finite sequence of coefficients $a_n$ of its Taylor series representation:
\begin{equation}\label{alg_ex_f1_ser}
f_0(z) = \sum_{n = 1}^{N_0} a_n z^n,
\end{equation}
with $n\in\{1,\ldots, N_0\}$. In the language of QFT modes, $f_0(z)$ is a `gapless' mode with $f_0(0) = 0$, so $a_0 = 0$. For this reason, a Taylor series representation around the origin mimics a result one would obtain from an EFT written in terms of a gradient expansion.

How does one proceed to reconstruct the spectrum? By using one of the standard convergence tests, the simplest thing one can do is to check that the series \eqref{alg_ex_f1_ser} has a finite radius of convergence, which is given by $|z_1|$. A priori, however, it is not clear how many critical points are obstructing its convergence at $\partial \mathbb{D}_{|z_1|}$. Assuming convergence along (most of) $\partial \mathbb{D}_{|z_1|}$, a concrete thing one can try is to investigate the behaviour of the series $f_1$ at the boundary by plotting its value or the value of its derivative. In many cases, this can serve as an excellent estimate for the number of branch points causing the divergence and, in turn, as a tool to pick the `correct' ansatz from the ones described in Section~\ref{sec:details}. 

In this case, we find only one critical point (the other critical point is indeed sufficiently far away), which means that we can use the results of the Darboux theorem discussed in Section~\ref{1cpt_section} (Case~\ref{AO2CP_versions_0}), in particular, in Subsection~\ref{sec:one_pt_alg}, where we outlined the procedure for the reconstruction from a finite sequence of known $a_n$.

\begin{figure}[ht!]
    \centering
    \includegraphics{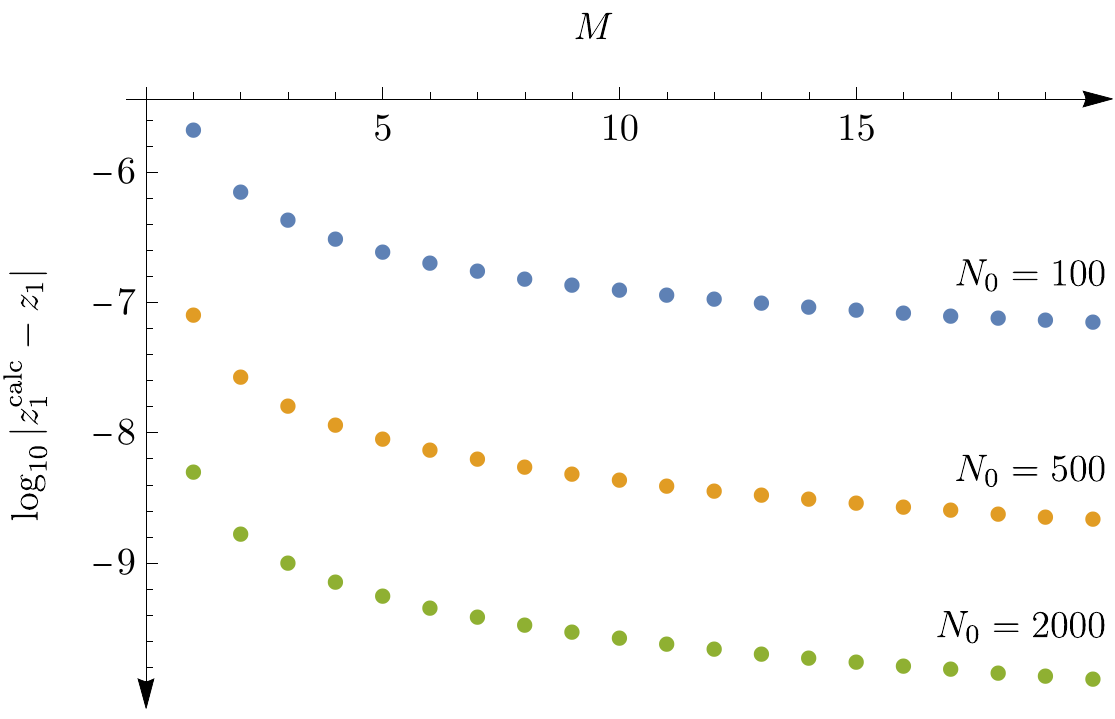}
    \caption{Error in the calculated branch point position plotted as a logarithm of $|z_1^{\text{calc}} - z_1|$ for different values of $M$, and a choice of three different values of $N_0 = 100$ (in blue), $N_0 = 500$ (in orange) and $N_0 = 2000$ (in green).}
    \label{AO1CP_numerics_a}
\end{figure}

We start by using the first part of the algorithm, which allows us to determine the location of the branch point. Namely, we construct the polynomials $X_{n\leq N_0}^{m\leq M}$ defined by the following recurrence relation from the coefficients $a_{n\leq N_0}$ (cf.~Eq.~\eqref{AO1CP_X_definition}):
\begin{equation}
\begin{aligned}
    X^0_n (\nu, z_1) &= a_n , \\
    X^{m+1}_n (\nu, z_1) &= X^m_n (\nu, z_1) - \frac{(n+\nu-2m-1)}{n z_1} X^m_{n-1} (\nu, z_1) , ~~ \text{for}~m \geq 0.
\end{aligned}
\end{equation}
By setting $X^m_{N_0} = 0$ and $X^m_{N_0 - 1} = 0$, for each  $m\leq M$ (cf.~Eq.~\eqref{bp_algo_HG_1}), and calculating the location and the order of the branch point, we obtain an approximation $z_1^{\text{calc}}$ of the critical point $z_1$. To show how well this procedure works for different $M$ and $N_0$, we compare $z_1^{\text{calc}}$ with the exact $z_1$ from Eq.~\eqref{P_alg_ex_cp} in Figure~\ref{AO1CP_numerics_a}. We see that for a fixed $N_0$, the precision slowly saturates with increasing $M$. If we desire results with higher precision, this cannot be achieved by simply increasing $M$. As expected, we have to increase the number of known coefficients $N_0$ as well. 

Having determined the branch point position to some desired precision, we then use the second part of the algorithm, which allows us to calculate all coefficients of the Puiseux expansion of $f_0$ and then $f_1$ around the branch point. To do this, we recursively compute the polynomials $Y_{\ell, n\leq N_0}^{m\leq M}$ defined in Eq.~\eqref{AO1CP_Y_definition}:
\begin{equation}
\begin{aligned}
Y^0_{\ell,n} (\nu, z_1) &= a_n , \\
Y^{m+1}_{\ell,n} (\nu, z_1) &= Y^m_{\ell,n} (\nu, z_1) - \frac{(n+\nu-2m-\ell - 2)}{n z_1} Y^m_{\ell, n-1} (\nu, z_1) , ~~ \text{for}~m \geq 0.
\end{aligned}
\end{equation}
Since we know the asymptotic form of the polynomials $Y_{\ell, n}^m$ (see~Eq.~\eqref{AO1CP_Y_asymptotic_form}), then, by evaluating the expression \eqref{coeffs_algo_HG_1}, we obtain 
\begin{equation}
r_\ell = \lim_{n\rightarrow\infty} \left[ \frac{(-1)^{\ell-\nu} n! z_1^{n+\nu-\ell}}{ m! (\nu-\ell)_{n-m}}Y_{\ell,n}^m - \sum_{k=0}^{\ell-1} \binom{m+\ell-k}{m} \frac{(-1)^{\ell-k}(\nu-k)_{n-m}r_k}{(\nu-\ell)_{n-m}z_1^{\ell-k}} \right],
\end{equation}
which gives half of the coefficients of the sought Puiseux series expansion: the odd $b_{2n+1}$. The remaining half, the even coefficients $b_{2n}$, follow from an analogous calculation for the auxiliary function $g(z)$, giving the coefficients $q_n$ from Eq.~\eqref{AO1CP_even_coeffs_expression}.

\begin{figure}[ht!]
    \centering
    \includegraphics{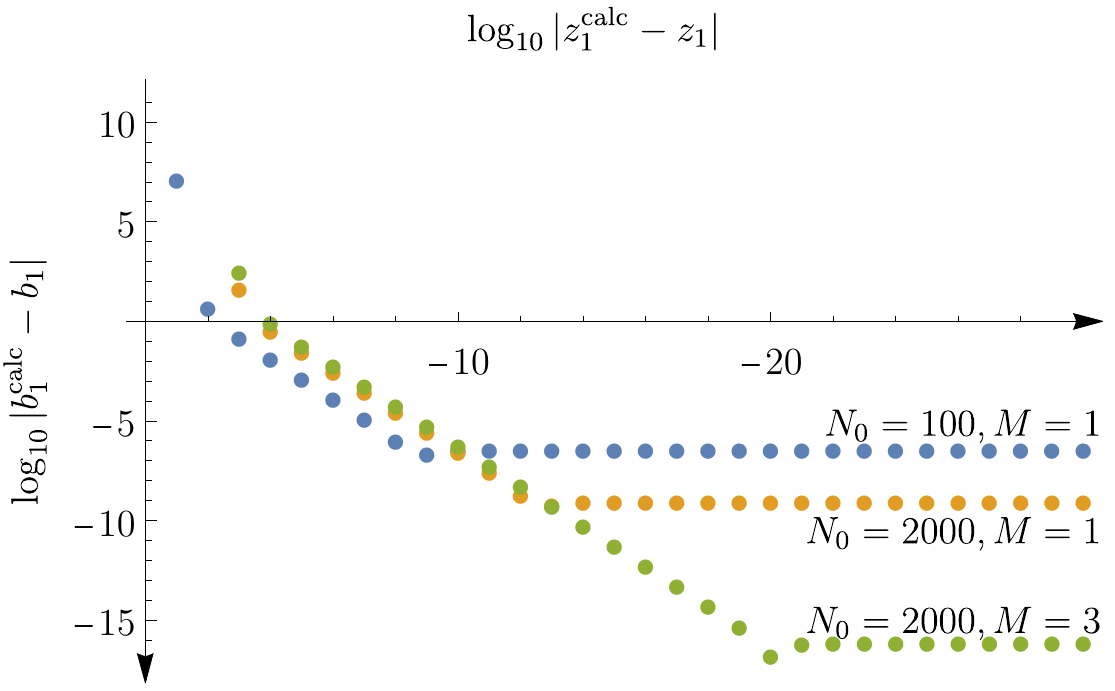}
    \caption{Error in the calculated coefficient $|b_1^{\text{calc}} - b_1|$ as a function of the error in the branch point position $|z_1^{\text{calc}} - z_1|$ for the following three choices of parameters: $N_0=100$, $M = 1$ (in blue), $N_0=2000$, $M=1$ (in orange) and $N_0 = 2000$, $M=3$ (in green).}
    \label{AO1CP_numerics_b}
\end{figure}

It is important to note that for the precision of the second step of the algorithm (determining $b_n$), both the number $N_0$ of the starting coefficients $a_n$ and the precision with which we determine $z_1^{\text{calc}}$ are essential. We demonstrate this by showing the error in the first calculated odd coefficient $b_1 = -i r_0$ (cf.~Eq.~\eqref{odd_and_even_coefficients}) versus the error in the calculated branch point $z_1^{\text{calc}}$ for three values of $N_0$ and $M$ used to calculate $b_1$. The results are shown in Figure~\ref{AO1CP_numerics_b}. For each $N_0$ and $M$ that we use to calculate the coefficient $b_1$, there exists a limiting precision of the branch point position above which $|b_1^{\text{calc}} - b_1|$ no longer depends on $|z_1^{\text{calc}} - z_1|$. Furthermore, the value of the limiting precision grows with increasing $N_0$ and $M$. Thus, when performing the calculations, it is best to try to keep the error in $z_1^{\text{calc}}$ below this particular limiting value so that the error in the subsequently calculated coefficients remains independent of it. 

Let us now also take a look at higher coefficients. In Table~\ref{AO1CP_bn_table_of_coefficients}, we show the first $5$ odd coefficients $r_n = b_{2n+1}e^{i \left(n+\frac{1}{2}\right)\pi}$ obtained by using different methods. The first column of results presents the `actual' (analytically obtained) values of $r_n$ that can be computed directly from the spectral curve (SC). In the second column, we show the coefficients calculated with the algorithm in Eq.~\eqref{coeffs_algo_HG_1}, using the (exact) analytic value of $z_1$. In the process, we use $N_0=2000$ starting coefficients and preform the recursion $M=5$ times. 

For the results in the third column, we instead use the calculated (reconstructed) value of the branch point $z_1^\text{calc}$, again using the algorithm in Eq.~\eqref{bp_algo_HG_1} with the same $N_0 = 2000$ and, now, $M=20$. By comparing the first three columns, we notice a rapid drop of precision of the calculated values of higher-order coefficients when $z_1$ is not determined precisely enough, i.e., when we use $z_1^\text{calc}$. In particular, we see that $r_1$ is determined to $5$ significant figures (as already determined from the two plots above), $r_2$ is precise only up to the first significant figure while $r_3$ is already dramatically wrong. We therefore conclude that with a fixed number of starting coefficients $N_0$, the error in the calculated position of $z_1$ indeed rapidly permeates the precision of calculated $r_n$.

Fortunately, in calculating $r_n$, this shortcoming can be somewhat remedied by employing an optimisation procedure to improve $z_1^\text{calc}$. We will call the $z_1$ calculated in this way the `optimised' branch point $z_1^\text{opt}$. The general idea behind the procedure is as follows. Since $|z_1|>1$, we know that the sequence of the coefficients $r_n$ we are trying to calculate is convergent, therefore, {\it Cauchy}. Hence, if the calculated coefficients start diverging, as they do in the third column of Table~\ref{AO1CP_bn_table_of_coefficients}, then we can attempt to correct the value of $z_1$ that we are using. Imagine now that we are calculating the first $n$ coefficients of the expansion. One thing we can do is define a quantity $F_n$ that measures the `Cauchyness' of the calculated coefficients, i.e.,
\begin{equation}
    F_n(r_{n-2},r_{n-1},r_n; z_1) = \frac{|r_{n-2}-r_{n-1}||r_{n-1}-r_{n}|}{\Big| |r_{n-2}-r_{n-1}| - |r_{n-1}-r_{n}| \Big|},
\end{equation}
where we chose for $F_n$ to depend on the last three of the calculated coefficients. Since the coefficients $r_n$ are calculated using a certain value of the branch point position, they implicitly all depend on the same $z_1$ that we used for their calculation. One can now simply minimise the value of $F_n$ while varying the value of the branch point position:
\begin{equation}
    \min_{z_1^\text{opt}\in\mathbb{C}} F_n(r_{n-2},r_{n-1},r_n; z_1^\text{opt}).
    \label{AO1CP_minimization}
\end{equation}
For the starting value of $z_1$, it is convenient to use the value of $z_1^\text{calc}$ calculated using Eq.~\eqref{bp_algo_HG_1}. Finding the value of $z_1^\text{opt}$ where the minimum is attained therefore allows us to determine the branch point position to precision much greater than that of $z_1^\text{calc}$. This is something that is not possible with the use of the algorithm \eqref{bp_algo_HG_1} alone. Nevertheless, by using the proposed optimisation (or a similar procedure), we do not need any additional information about the solution $f_0$, only the initial $N_0$ coefficients. The results of such a computation for the first $5$ odd coefficients, where we use Eq.~\eqref{AO1CP_minimization} with $n=5$, are included in the last column of Table~\ref{AO1CP_bn_table_of_coefficients}. We note that by doing this, we nearly doubled the precision of the calculated branch point and managed to reconstruct $3$ instead of $1$ coefficient with the precision of $5$ significant figures or more. 

Since we want this to serve as a proof of principle, we will not explore such optimisations any further and we will hereon for simplicity use the analytically computed value of the branch point $z_1$ to demonstrate the rest of the calculation. Another reason to continue with $z_1$ is that the success in calculating $z_1^\text{calc}$ and its importance for the calculation of coefficients appear to be rather case-specific. As we will see in the holographic example in Section~\ref{sec:M2}, no optimisation of the branch point location will be necessary to obtain `sufficiently good' results for the purposes of this paper.

\begin{table}[ht!]
\centering
\begin{tabular}{|c||c|c|c|c||} 
 \hline
 $n$  & $r_n$ (from SC) & $r_n^\text{calc}$ (with $z_1$) & $r_n^\text{calc}$ (with $z_1^\text{calc}$) & $r_n^\text{calc}$ (with $z_1^\text{opt}$) \\ [0.5ex] 
 \hline\hline
1 & $0.6147881530$ & $0.6147881530$ & $0.6147866244$ & $0.6147881530$ \\
2 & $-0.0207471500$ & $-0.0207471500$ & $-0.0127043545$ & $-0.0207471466$ \\
3 & $0.0032346953$ & $0.0032346953$ & $-13.532036923$ & $0.0032288977$ \\
4 & $-0.0006892719$ & $-0.0006892719$ & $9147.5180459$ & $0.0032288977$ \\
5 & $0.0001694157$ & $0.0001694150$ & $-2.5631597\times 10^{6}$ & $-1.0977124652$ \\
 \hline
 \hline
  \makecell{branch point \\ precision} & / & $-\infty$ & $-9.882$ & $-16.250$ \\
 \hline
\end{tabular}
\caption{The first $5$ coefficients $r_n$ determining the odd coefficients $b_{2n+1}$ of the expansion of $f_1(z)$ around $z_1$. The first column displays the `exact' values of $r_n$ computed directly (analytically) from the spectral curve (SC). The remaining three columns show the `reconstructed' values of $r_n$ (using the algorithm in Eq.~\eqref{coeffs_algo_HG_1}), computed by using the location of the critical point $z_1$ obtained in different ways: exact $z_1$, reconstructed $z_1^\text{calc}$ from the Darboux theorem and optimised $z_1^\text{opt}$ (cf.~Eq.~\eqref{AO1CP_minimization}), respectively. In the last line of the table, we state the error $\log_{10}|z_1^\text{calc}-z_1|$ of the branch point used in each respective column.}
\label{AO1CP_bn_table_of_coefficients}
\end{table}

If we continue with the analytically known value of the branch point $z_1$, it is enough to use a smaller number of initial coefficients. For example, we can take $N_0 = 200$ and calculate the coefficients $b_n$ by performing the recursion $M=100$ times. We find this to be sufficient to calculate $80$ coefficients ($40$ even and $40$ odd as designated in \eqref{odd_and_even_coefficients}) of the Puiseux expansion of $f_0$ around $z_1$. Since the expansions of $f_0(z)$ and $f_1(z)$ around $z_1$ are related by Eq.~\eqref{f_ser_both}, this means that we have obtained the series representation of the second solution $f_1(z)$ around $z_1$:
\begin{equation}
f_1(z) = \sum_{n=0}^{N_1-1} b_n (z-z_1)^{n/2},
\label{ao1cp_example_1st_cp_expansion}
\end{equation}
with $N_1 = 80$. The solution $f_1$ is the second Riemann sheet of $f_0$ at $z_1$. In the language of physical modes, this is the first `gapped mode' with $f_1(z=0) \neq 0 $. In fact, a concrete and physically motivated question that allows us to check these results is the calculation of the value of the gap $f_1(z=0)$ (see e.g.~Figure~\ref{fig:AO1CP_made_up_example}). By simply evaluating the series \eqref{ao1cp_example_1st_cp_expansion} at $z=0$, we obtain the correct value of $f_1(0) = 1$ to $30$ significant figures.

We can now further investigate the structure of the full Riemann surface (of the spectrum) by continuing with the same procedure, using the calculated coefficients $b_n$. The next natural step is the re-expansion of $f_1$ around the critical point that obstructs the convergence of the series \eqref{ao1cp_example_1st_cp_expansion}. This point is $z_2$ and we now again show how to determine its position and the Puiseux coefficients of the expansion around it using only the coefficients $b_n$ just calculated.

A small complication in performing the second step of the re-expansion (going from $z_1$ to $z_2$) arises due to the assumption of regularity of $r(z)$ and $q(z)$ in the ansatz \eqref{f_crit_exp}. This assumption is no longer true as the starting series \eqref{ao1cp_example_1st_cp_expansion} is now a Puiseux series of order $p=2$ that is expanded around a critical point. What we do to re-expand the series around $z_2$ by means of the Darboux theorem is to change the ansatz in Eq.~\eqref{f_crit_exp} and treat the odd and the even coefficients of the starting expansion independently. We describe this `critical origin' algorithm in detail in Appendix~\ref{app11}.\footnote{An alternative way of dealing with the critical origin of the expansion is to use the 'unwinding' method discussed in the Appendix~\ref{app12}.}

From the $N_1=80$ coefficients of the Puiseux expansion around the point $z_1$, we calculate $N_2=10$ coefficients of the Puiseux expansion of $f_1$ around $z_2$. Then, by again using Eq.~\eqref{f_ser_both}, we obtain the third solution $f_2$ with the coefficients $c_n$:
\begin{equation}
    f_2(z) = \sum_{n=0}^{N_2-1} c_n (z-z_2)^{n/2}.
    \label{ao1cp_example_2nd_cp_expansion}
\end{equation}
The calculated coefficients $c_n$ are compared to their analytical values obtained from the spectral curve (SC) in Table~\ref{AO1CP_table_of_coefficients}. We can also calculate the `gap' $f_2(0)$ by summing up the truncated series expansion \eqref{ao1cp_example_2nd_cp_expansion} of the $N_2$ terms and evaluating it in $z=0$. The value we obtain is $f_2(0) = 2.9996208363$, which is in good agreement with the true value of $f_2(0) = 3$.

\begin{table}[ht!]
\centering
\begin{tabular}{|c||c|c||} 
 \hline
 $n$  & $c_n$ (from SC) & $c_n$ (reconstucted) \\ [0.5ex] 
 \hline\hline
 0 & $2.2152504370$ & $2.2152504369$ \\ 
 1  & $0.6147881530$ & $0.6147881501$ \\
 2  & $-0.0714285714$ & $-0.0714285970$\\
 3  & $0.0207471500$ & $0.0207469757$ \\
 4 & $-0.0077135607$ & $-0.0077142868$ \\
 5  & $0.0032346953$ & $0.0032319490$ \\
 6  & $-0.0014577259$ & $-0.0014647362$\\ 
 7  & $0.0006892719$ & $0.0006719018$\\ 
 8  & $-0.0003373277$ & $-0.0003676770$\\ 
 9  & $0.0001694157$ & $0.0001164412$\\ %[1ex] 
 \hline 
\end{tabular}
\caption{Comparison between the $N_2=10$ calculated coefficients $c_n$ of the expansion of $f_2(z)$ around $z_2$. The two columns show the analytically obtained values from the spectral curve (SC) and the reconstructed values using the Darboux theorem.}
\label{AO1CP_table_of_coefficients}
\end{table}

\begin{figure}[ht!]
    \centering
    \includegraphics{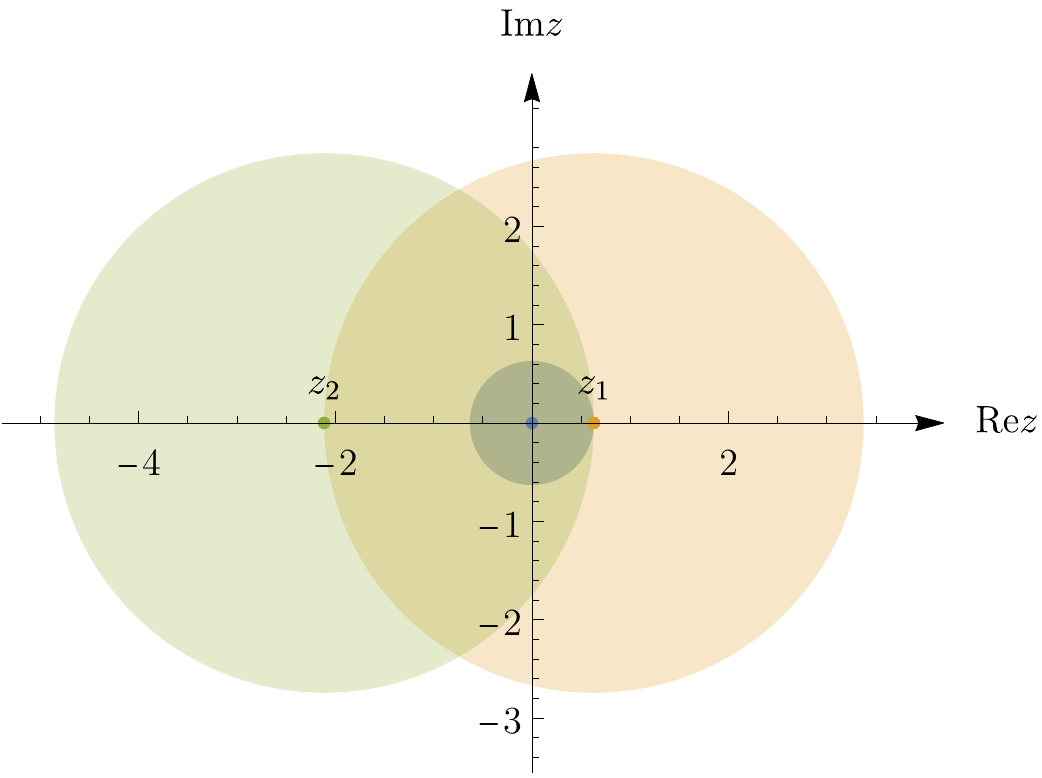}
    \caption{Locations of the points $z=0$, $z_1$ and $z_2$ and the disks of convergence of the series expansions of $f_0$ (Eq.~\eqref{alg_ex_f1_ser}), $f_1$ (Eq.~\eqref{ao1cp_example_1st_cp_expansion}) and $f_2$ (Eq.~\eqref{ao1cp_example_2nd_cp_expansion}) around $z=0$, $z_1$ and $z_2$, respectively.}
    \label{AO1CP_critical_points}
\end{figure}

With the knowledge of the three expansions \eqref{alg_ex_f1_ser}, \eqref{ao1cp_example_1st_cp_expansion} and \eqref{ao1cp_example_2nd_cp_expansion} around the three points $z=0$, $z=z_1$ and $z=z_2$, respectively, we have, in regions, reconstructed all branches of the Riemann surface of solutions of the equation $P(f,z)=0$ in \eqref{P_alg_ex}. We show the two critical points and the three disks of convergence of $f_0$, $f_1$ and $f_2$ in Figure~\ref{AO1CP_critical_points}. Importantly, {\it all} points that are outside of the respective disks of convergence of each of the three series expansions can then be reached through analytic continuation of the three series (without the need for any additional extensions to other Riemann sheets).\footnote{Concretely, one can use for example a M\"{o}bius transformation in order to transform the disk of convergence into a half-plane of convergence or one of many other methods of analytic continuation.}

In conclusion, for the simple toy model of an algebraic cubic complex spectral curve, we were able to `reconstruct' all branches of the entire Riemann surface just from the knowledge of the coefficients $a_n$ of the Taylor expansion around the origin. Given the convergent nature of the procedure, we expect that as $n\to\infty$, our algorithm can succeed with the reconstruction to an arbitrarily high precision. 

\subsection{Transverse momentum and diffusion of M2 branes}\label{sec:M2}

In this section, we turn our attention to the main example studied in this work. This is the 3$d$ large-$N$ CFT that describes the dynamics of a large-$N$ stack of M2 branes of which the energy-momentum sector is holographically dual to gravitational fluctuations in the AdS${}_4$-Schwarzschild black brane geometry. In terms of the ingoing Eddington-Finkelstein coordinates $(v,r,x,y)$, the linearised metric perturbations can be Fourier decomposed along the flat 3$d$ spacetime as $\sim e^{- i \omega v + iqx} h_{\mu\nu}(r)$. Here, we will only study the transverse fluctuations (with respect to the wavevector pointing along the $x$ axis). The gapless hydrodynamic mode in the spectrum is a diffusive mode, which was first studied by Herzog in Ref.~\cite{Herzog:2002fn}.

To analyse the spectrum of transverse momentum fluctuations, we use standard holographic techniques (see~Ref.~\cite{Kovtun:2005ev}). We look for the equation of motion of the gauge-invariant mode $Z(r) \equiv (\omega h_{xy}(r) + q h_{ty}(r))/r^2$, where the radial coordinate runs from the event horizon at $r=r_0$ to the boundary at $r\to\infty$. The Hawking temperature is $T = 3 r_0 / 4 \pi $. Instead of using $r$, we will work with a radial coordinate $u$, defined as $u = (r_0/r)^{3/2}$, with the horizon now located at $u=1$ and the boundary at $u=0$. The equation of motion for $Z(u)$ is then given by
\begin{equation}\label{Z_eom}
    \begin{aligned}
        \partial_u^2 Z(u) &+ \frac{\wfr^2(1+u^2-2iu^{2/3}\wfr)-z(u^2-1)(u^2+2iu^{2/3}\wfr-1)}{u(u^2-1)(\wfr^2+(u^2-1)z)}\partial_u Z(u) \\ &+ \frac{3z^2u^{2/3}(u^2-1)+2i\wfr z(5u^2-2)+3zu^{2/3}\wfr^2+4i\wfr^3}{3u^{4/3}(u^2-1)(\wfr^2+(u^2-1)z)}Z(u) = 0,
    \end{aligned}
\end{equation}
where we have defined the following dimensionless quantities: $\wfr \equiv \omega / (2\pi T)$ and $z \equiv \qfr^2$, where $\qfr \equiv q / (2 \pi T) $. The quasinormal mode spectrum of $Z(u)$ (with $Z(u)$ regular at $u=1$) evaluated at the AdS boundary $u=0$ gives the spectrum of the dual retarded correlator \cite{Son:2002sd}. In the language of Refs.~\cite{Grozdanov:2019kge,Grozdanov:2019uhi}, $Z(u)$ evaluated at $u=0$ is (proportional to) the complex spectral curve $Z(u=0) = P(\wfr, z)$. To find $P(\wfr,z)$, we solve the equation \eqref{Z_eom} for $Z(u)$ in a Frobenius expansion around the horizon in powers of $(1-u)$ to order $30$. This numerical procedure turns the problem into an analysis of an algebraic spectral curve of which the spectrum, we assume, converges to the `actual' full quasinormal spectrum as the order of the curve is increased. 

\begin{figure}[ht!]
    \centering
    \includegraphics{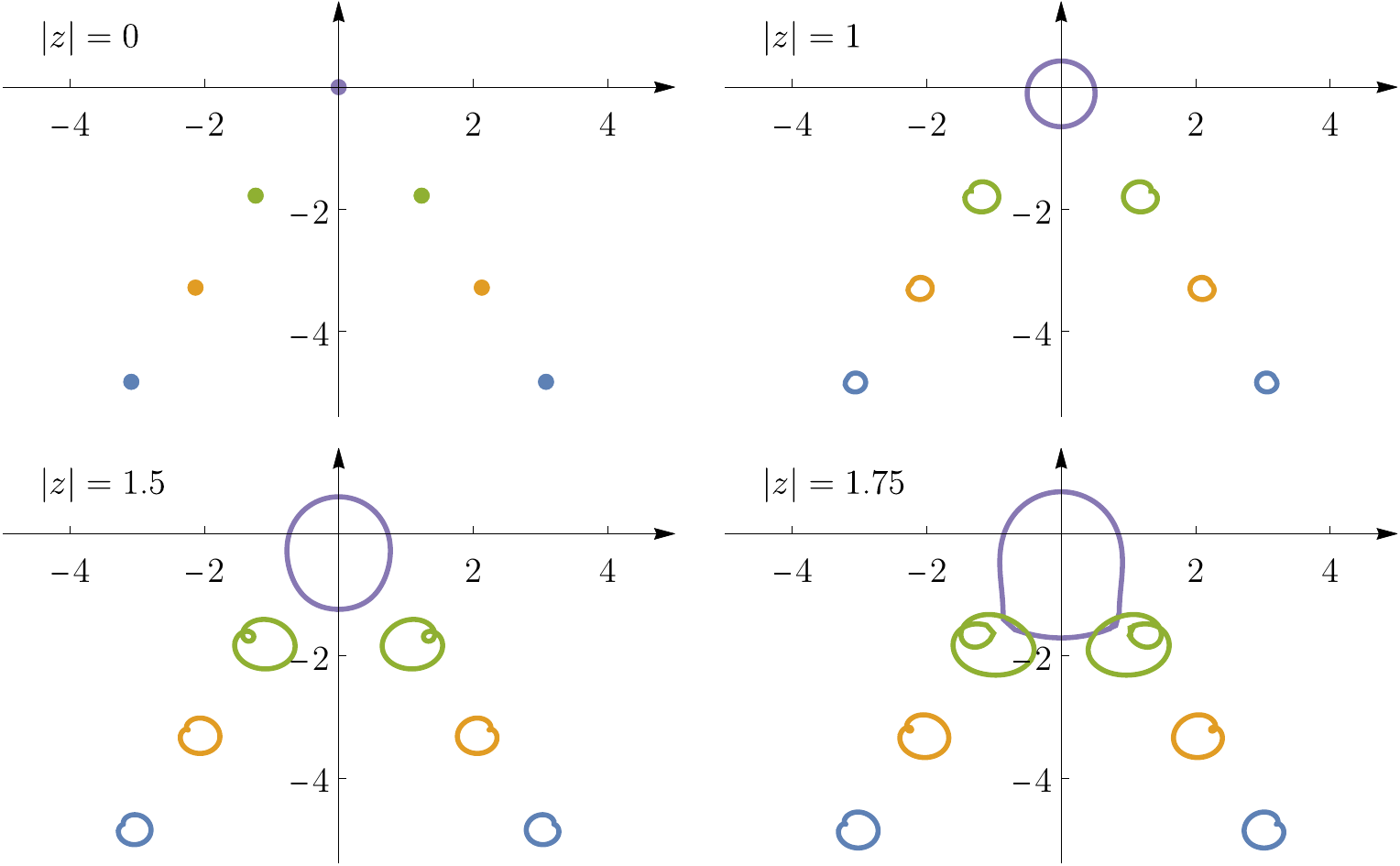}
    \caption{Quasinormal mode solutions of the spectral curve equation $P(\wfr,z)=0$ in the complex $\wfr$-plane at fixed $|z|$ with varying $\text{Arg}[z] \in [0,2\pi]$.}
    \label{fig:AdS4_before_mode_collisions}
\end{figure}

We now analyse the spectral curve $P(\wfr,z)$ in some detail. By solving the equation $P(\wfr,z) = 0$ numerically, we first show the spectrum for $4$ different `small' choices of $|z|$ (before any level-crossing occurs) in Figure~\ref{fig:AdS4_before_mode_collisions} (see Refs.~\cite{Grozdanov:2019kge,Grozdanov:2019uhi} for a detailed discussion of how such results are obtained). Besides the hydrodynamic diffusive mode $\wfr_0(z)$ (plotted in purple in Figure~\ref{fig:AdS4_before_mode_collisions}), we will mainly focus on the lowest two gapped modes $\wfr_1(z)$ (plotted in green) and $\wfr_2(z)$ (plotted in orange), which belong to the so-called `Christmas tree' part of the spectrum.

To find the desired series expansion of the hydrodynamic diffusive mode $\wfr_0(z)$, we use the equation $P(\wfr,z) = 0$ and directly calculate the first $N_0 = 300$ coefficients $a_n$ of the hydrodynamic series
\begin{equation}\label{wfr_1}
    \wfr_0(z) = \sum_{n = 1}^{N_0} a_n z^n,
\end{equation}
where all $a_n$ are imaginary. Note that since $\wfr_0(z)$ is gapless (i.e., $\wfr_0(z = 0) = 0$), $a_0 = 0$. Note also that $a_1 = - 2 \pi T i D$, where $D$ is the momentum diffusivity that can be expressed in terms of shear viscosity $\eta$, entropy density $s$ and temperature. In this theory, at infinitely strong coupling, $D = \eta/(s T) = 1/(4\pi T)$ \cite{Herzog:2002fn,Kovtun:2004de}.

\begin{figure}[ht!]
    \centering
    \includegraphics{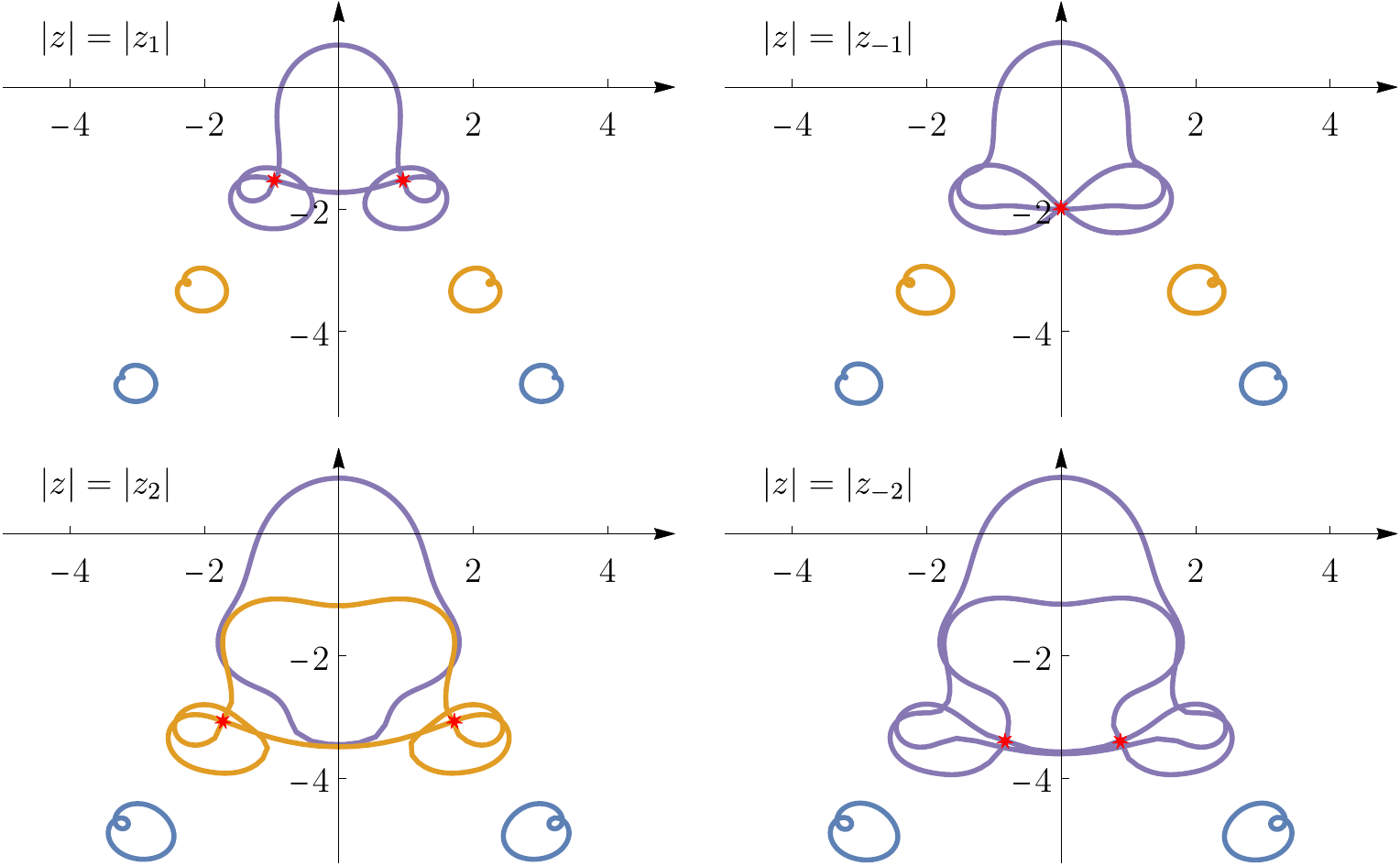}
    \caption{Quasinormal mode solutions of the spectral curve equation $P(\wfr,z)=0$ in the complex $\wfr$-plane at fixed $|z|$ with varying $\text{Arg}[z] \in [0,2\pi]$. The values of $|z|$ are chosen precisely at the four lowest level-crossing critical points $z_1$, $z_{-1}$, $z_2$ and $z_{-2}$ to show the pattern of collisions. The points where the collisions between the modes occur are depicted with red stars.}
    \label{fig:AdS4_mode_collisions}
\end{figure}

\begin{figure}[ht!]
    \centering
    \includegraphics[width=0.75\textwidth]{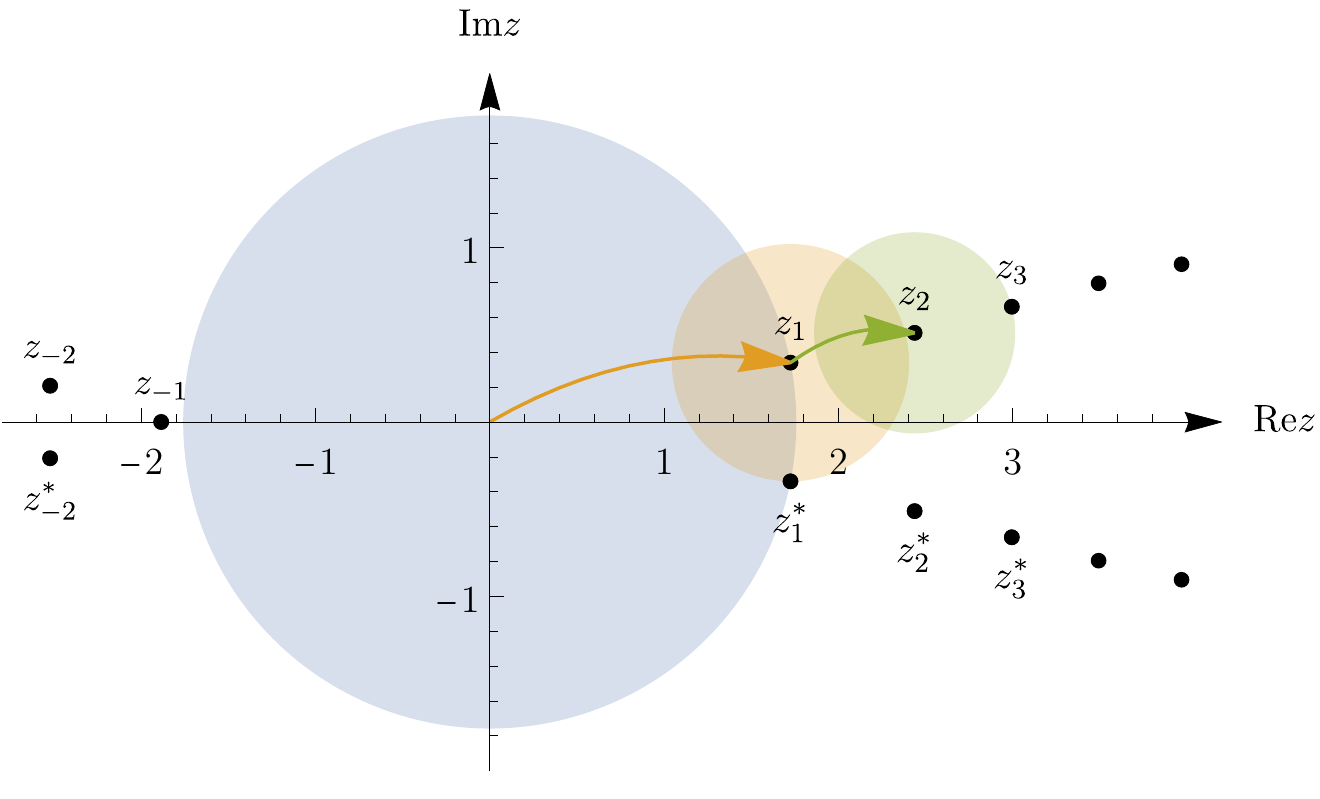}
    \caption{The locations of the first few critical points of $P(\wfr,z)$ in the transverse momentum (shear) channel of the AdS${}_4$-Schwarzschild black brane plotted in the complex $z$-plane. The coloured disks represent the disks of convergence of the three expansions of $\wfr_0$, $\wfr_1$ and $\wfr_2$ around the points $z=0$, $z_1$ and $z_2$, respectively. The arrows indicate the re-expansion steps undertaken to reconstruct the spectrum.}
    \label{fig:AdS4_critical_points}
\end{figure}

The series expansion \eqref{wfr_1} converges inside a disk of radius $|z_1|$ set by the (level-crossing) critical point $z_1$ of the spectral curve $P(\wfr, z)=0$ that is closest to the origin. We plot the quasinormal spectrum at the first four level-crossing occurrences in Figure~\ref{fig:AdS4_mode_collisions}. The approximate numerical values of the first few critical points (see Eq.~\eqref{critical_point}) are:
\begin{equation}
\begin{aligned}
    z_1 &= 1.72557 + 0.340429 i ,&\quad \wfr(z_1) &= 0.925957 - 1.524804 i, \\
    z_2 &= 2.43776 + 0.511447 i ,&\quad  \wfr(z_2) &= 1.661421 - 3.084745 i, \\
    z_3 &= 2.99533 + 0.661373 i ,&\quad  \wfr(z_3) &= 2.494940 - 4.672911 i , \\
    z_{-2} &= -2.52238 + 0.208041 i, &\quad \wfr(z_{-2}) &= -0.836760 - 3.414055 i, \\
    z_{-1} &= -1.88524, &\quad \wfr(z_{-1}) &= -1.977860, \\
\end{aligned}
\end{equation}
where the negative subscripts indicate the fact that the point lies in the negative $\text{Re} z$ half-plane of the complex $z$-plane. We depict the positions of the lowest critical points in Figure~\ref{fig:AdS4_critical_points}. By comparing the present case to the algebraic curve example studied in Subsection~\ref{ao1cp_example}, one immediately notices that there is now a larger number of critical points and that, more importantly, some points are located `very close' to each other. While this does not prevent a reconstruction, it does make it more difficult as different level-crossing critical points (of their respective $\wfr_n$) present obstructions to certain convenient re-expansions of the series. To show this, besides the critical points, in the Figure~\ref{fig:AdS4_critical_points}, we also plot three disks of convergence of the expansions $\wfr_0$ (the hydrodynamic diffusive mode) and the first two gapped modes $\wfr_1$ and $\wfr_2$ around $z=0$, $z_1$ and $z_2$, respectively. The blue disk is the (original) disk of convergence of the hydrodynamic diffusive Taylor series while the orange and green disks represent the disks of convergence of the Puiseux series of the first two gapped modes around $z_1$ and $z_2$, respectively. The reconstruction using the Darboux theorem will follow the arrows from the origin to the first and the second gapped modes. 

\begin{figure}[h!]
    \centering
    \includegraphics[scale=0.9]{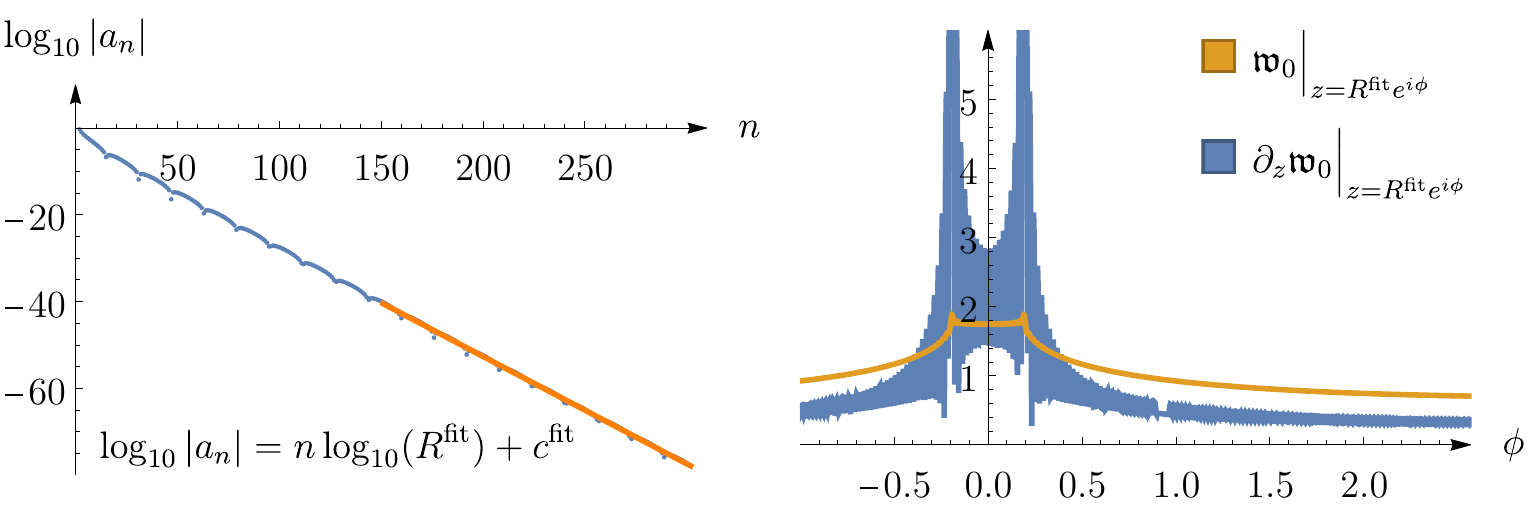}
    \caption{A non-rigorous `exploratory'  approach to determining the nature of the obstruction to the convergence of the series \eqref{wfr_1}. \textbf{Left:} Calculation of the radius of convergence $R$ by use of a simple root test. For the linear fit, we only use the second half of all calculated coefficients. \textbf{Right:} Plot of the series representations of $\wfr(z)$ and $\partial_z \wfr(z)$ for $z=R^\text{fit} e^{i\phi}$, where $\phi\in[0,2\pi]$. In this case, one can correctly identify the locations of diverging derivatives with positions of the branch points $z_1$ and $z_1^*$.}
    \label{fig:AdS4_root_and_plot}
\end{figure}

After this brief introduction to the features of the holographic spectrum at hand, we proceed with the reconstruction by using the theorems of Darboux and Puiseux. For comparison of our results with a reconstruction that uses the method of Pad\'{e} approximants, see Appendix~\ref{app3}. The first step is to re-expand the hydrodynamic series \eqref{wfr_1} around the first critical point $z_1$ where the hydrodynamic mode collides with the gapped mode $\wfr_1$. This step is depicted with the orange arrow in Figure~\ref{fig:AdS4_critical_points}. As discussed, in general, we do not know how many critical points are obstructing the convergence of the starting series. Therefore, it is sometimes useful, and sometimes effective, to employ certain simple methods that explore the behaviour of \eqref{wfr_1} at the boundary of the disk of convergence. In this particular case, we start by applying the root convergence test to the coefficients $a_n$ to find the approximate radius of convergence. In the left panel of Figure~\ref{fig:AdS4_root_and_plot}, we display the logarithms of the coefficients along with a linear fit, which gives $R^\text{fit}=1.76937$ to $2$ significant figures (the `actual' radius of convergence is $R= 1.75883$). We then set $z=R^\text{fit}e^{i\phi}$ and plot $\wfr(z)$ and $\partial_z \wfr(z)$ as functions of $\phi\in[0,2\pi]$ to observe their behaviour at the boundary. In this way, we can get a sense of the number of critical points that are located on or near the boundary of the convergence disk. This is shown in the right panel of Figure~\ref{fig:AdS4_root_and_plot}.

\begin{figure}[ht!]
    \centering
    \includegraphics[scale=0.9]{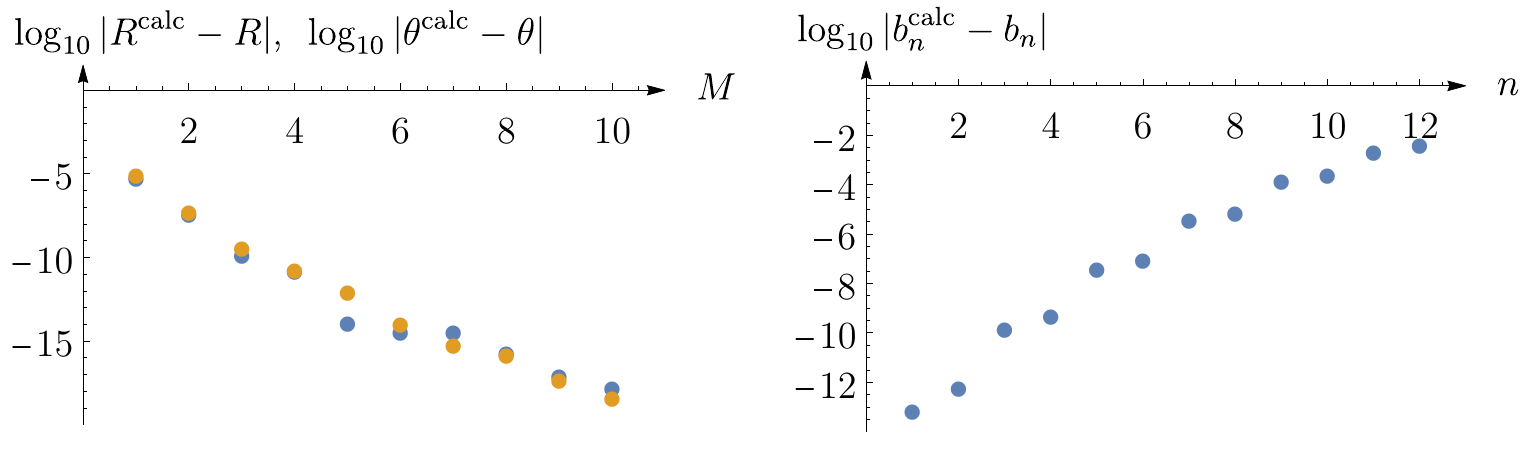}
    \caption{Plot of the errors of our results after the use of the Case~\ref{AO2CP_versions_a} algorithms in Eqs.~\eqref{ao2cp_bp_algorithm} and \eqref{ao2cp_coeffs_algorithm}. \textbf{Left:} Error in the calculated values of $R$ (in blue) and $\theta$ (in orange) versus $M$ (for $N_0=300$). \textbf{Right:} Error in the first $12$ coefficients $b_n$ of the expansion of $\wfr_1(z)$ around $z=z_1$.}
    \label{fig:AdS4_bp_coeffs_numerics}
\end{figure}

We observe that, likely, we are dealing with a case of two critical point causing the divergence of $\partial_z \wfr(z)$, and thereby, the divergence of the hydrodynamic series \eqref{wfr_1}. Moreover, the two locations (as required by the symmetries of the spectrum) are each other's complex conjugates, which means that it is sensible to proceed by using the Case~\ref{AO2CP_versions_a} of the reconstruction algorithm discussed in Section~\ref{sec:details}. If this choice were for some reason incorrect, we would immediately notice this by the lack of convergence of the algorithm, which could then be remedied by a different ansatz. In the language of the relevant Section~\ref{sec:2pt_a}, we denote the modulus and the argument of the obstructing branch point $z_1$ by $R=|z_1|$ and $\theta=\text{Arg}[z_1]$. We now use the Darboux theorem to determine $z_1$ to a much greater precision with the use of the algorithm in Eq.~\eqref{ao2cp_bp_algorithm}. In particular,  we take $N_0=300$ initial coefficients and perform the recursion $M=10$ times. The errors in the calculated $R^\text{calc}$ and $\theta^\text{calc}$ are shown in the left panel of Figure~\ref{fig:AdS4_bp_coeffs_numerics}. What we see is that by using this procedure, we are able to determine the position of the branch point to approximately  $18$ significant figures.

Next, we calculate the Puiseux series expansion of $\wfr_0(z)$ around one of the two closest critical points, $z_1$. By using the algorithm in Eq.~\eqref{ao2cp_coeffs_algorithm} with $M=1$, we are able to obtain the first $N_1=12$ coefficients of the series. The coefficients themselves are shown in Table~\ref{AdS4_table_of_coefficients_1}, while their errors are plotted in the right panel of Figure~\ref{fig:AdS4_bp_coeffs_numerics}. Although the results are satisfactory, the algorithm in this case does not converge as fast as in the case of a single critical point in Section~\eqref{sec:one_pt_alg} that was used for the reconstruction of the cubic algebraic curve \eqref{ao1cp_example} of Section~\ref{ao1cp_example}. However, it is also important to note that the success of the calculation of the coefficients has significantly less dependence on the precision with which the branch point location is calculated than in the previous section. 

Having obtained the Puiseux series expansion of $\wfr_0(z)$ around the critical point $z=z_1$, we now also know the series expansion of the second Riemann sheet, the gapped mode $\wfr_1(z)$ around $z=z_1$. As before, the two are related by Eq.~\eqref{f_ser_both} and therefore,
\begin{equation}
    \wfr_1(z) = \sum_{n=0}^{N_1-1} b_n (z-z_1)^{n/2},
    \label{ads4_w1_exp}
\end{equation}
where, recall, $N_1=12$. 
\begin{table}[ht!]
\centering
\begin{tabular}{|c||c|c||} 
 \hline
 $n$ & $b_n$ (from SC) & $b_n$ (reconstructed) \\ [0.5ex] 
 \hline\hline
0 & $0.9259568405-1.5248039730i$ & $0.9259568405-1.5248039730i$ \\
1 & $0.5694243295+0.8593408725i$ & $0.5694243295+0.8593408725i$ \\
2 & $0.2465534194-0.6564234393i$ & $0.2465534194-0.6564234391i$ \\
3 & $-0.2017505274+0.4158971322i$ & $-0.2017505276+0.4158971316i$ \\
4 & $0.2063443343-0.2276803807i$ & $0.2063443322-0.2276804330i$ \\
5 & $-0.1815319274+0.1089739687i$ & $-0.1815318686+0.1089740741i$ \\
6 & $0.1176312182-0.0460907198i$ & $0.1176321889-0.0460860162i$ \\
7 & $-0.0445357280+0.0324548520i$ & $-0.0445414084+0.0324475608i$ \\
8 & $-0.0067265055-0.0593218510i$ & $-0.0067939136-0.0594884583i$ \\
9 & $0.0191994441+0.1038324422i$ & $0.0194382654+0.1040542804i$ \\
10 & $0.0053380839-0.1353137209i$ & $0.0069628396-0.1329437874i$ \\
11 & $-0.0460909505+0.1306640455i$ & $-0.0507261321+0.1276808356i$ \\
 \hline 
\end{tabular}
\caption{Comparison between the $N_1=12$ coefficients $b_n$ form the series expansion of $\wfr_1(z)$ calculated directly from the spectral curve (SC) and those reconstructed by using the Darboux theorem.}
\label{AdS4_table_of_coefficients_1}
\end{table}
A clear physically motivated question is again the calculation of the gap  $\wfr_1(z=0)$ of the first excited mode $\wfr_1(z)$. This is now not completely straightforward because, as can be checked, the series \eqref{ads4_w1_exp} has a radius of convergence, which does not extend to $z=0$. This can be seen from Figure~\ref{fig:AdS4_critical_points}, where the disk of convergence is shaded with orange colour. In fact, the critical points $z_1^*$, $z_2$ and $z_2^*$ are all closer to $z_1$ than the origin. This means that we are required to perform an analytic continuation within the same Riemann sheet to evaluate $\wfr_1(0)$. While this can be done by any one of numerous methods, we find that a simple use of a Pad\'e approximant around $z=z_1$ provides a rather effective analytic continuation. Using the approximant of order $[N_1/2, N_1/2]$, we determine the gap $\wfr_1^\text{calc}(0)$, which can be compared with the value of $\wfr_1(0)$ computed directly from the spectral curve. Their values are:
\begin{align}
    \wfr_1^\text{calc}(0)&=1.23506 - 1.76338 i, \\
    \wfr_1(0) &= 1.23455 - 1.77586 i.
\end{align}
The results agree to approximately two significant figures. Furthermore, with an analytic continuation that can access the dispersion relation of the first gapped mode at $z = 0$, we could at this point, as in Ref.~\cite{Abbasi:2020xli}, also look for the structure of $\wfr_1(z)$ expanded in terms of a gradient expansion in powers of $z = q^2$. Concretely, we could find the coefficients $a^{(1)}_n$ of the series $\wfr_1(z) = \wfr_1(0) - i \sum_{n=1}^\infty a^{(1)}_{n} z^{n}$. 

To carry out the next step of the reconstruction (depicted by the green arrow in Figure~\ref{fig:AdS4_critical_points}) with `reasonable' precision, we need more than the calculated $N_1=12$ coefficients obtained from the first step. In principle, this is not a problem. However, due to computational limitations, and since the purpose of this work is to show the general power of reconstructions and our concrete algorithm at work, we, at that point, choose to recompute the coefficients of the Puiseux series \eqref{ads4_w1_exp} directly from the spectral curve $P(\wfr, z)=0$. This drastically improves the precision of the second step. We do this by using the Newton-Puiseux (polygon) method, discussed, for example, in Ref.~\cite{Grozdanov:2019uhi}. Concretely, we compute $N_1 = 300$ coefficients and continue to perform the second step of the reconstruction by following the same procedure as in step one. 

\begin{figure}[ht!]
    \centering
    \includegraphics[scale=0.9]{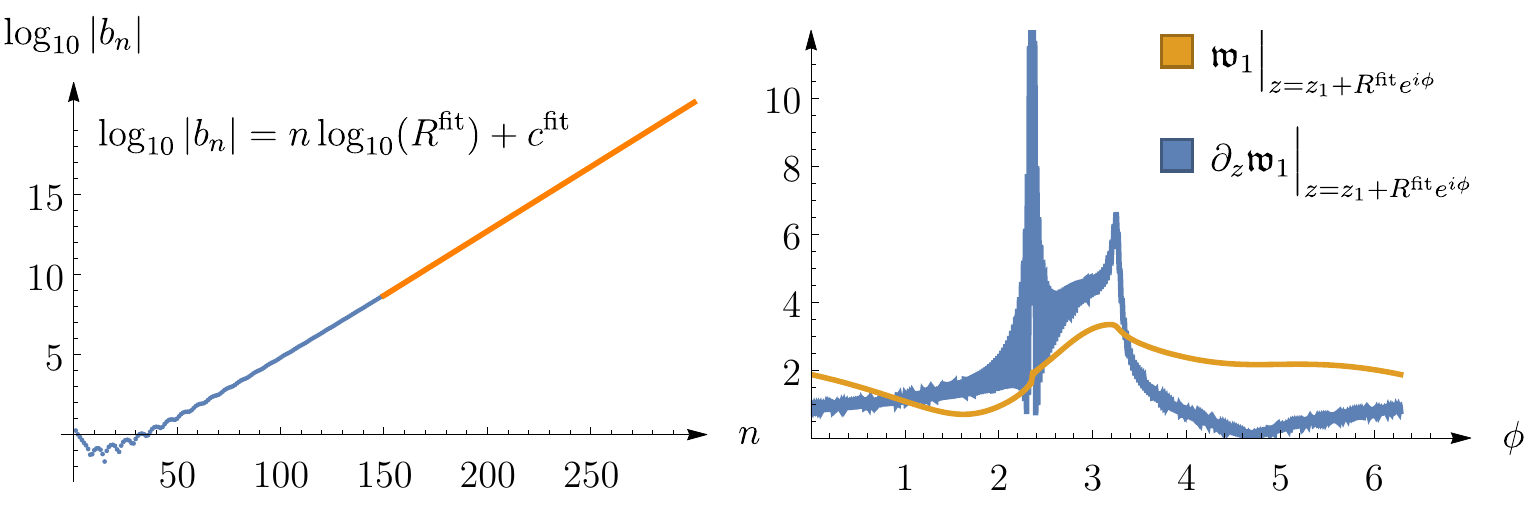}
    \caption{Plots determining the nature of the obstruction to convergence of the series representation \eqref{ads4_w1_exp} of $\wfr_1(z)$. \textbf{Left:} Calculation of the radius of convergence $R^\text{fit}$ of \eqref{ads4_w1_exp} by the root test method. For the linear fit, we again use only the second half of all calculated coefficients. \textbf{Right:} Plot of the series representations of $\wfr_1(z)$ and $\partial_z \wfr_1(z)$, for $z=z_1 + R^\text{fit} e^{i\phi}$, where $\phi\in[0,2\pi]$. One can identify the local maxima of the derivative with the two closest branch points. The larger and more prominent (`divergent') peak corresponds to $z_1^*$ and the smaller to $z_2$. The branch point $z_2$ is located outside the convergence disk of the series.}
    \label{fig:AdS4_root_and_plot_2}
\end{figure}

In Figure~\ref{fig:AdS4_root_and_plot_2}, we show the results of the convergence test of the series \eqref{ads4_w1_exp} and a simple, naive analysis of the location of critical point at the edge of the convergence disk. Importantly, we see only one `actual' critical point obstructing convergence. However, what we can further infer is that the pattern of $\partial_z \wfr_1$ signals the presence of another critical point near the disk of convergence. One of these (the closest one) is the conjugate point $z_1^*$ while the second point is a new critical point $z_2$ around which we actually wish to re-expand the series. 

In cases where, strictly, only one critical point obstructs convergence but there is another point near the boundary of the disk of convergence, the most numerically efficient way to proceed is to use the Case~\ref{AO2CP_versions_c} algorithm of the Darboux theorem with two (independent) branch points (using \eqref{AO2CP_NC_NE}), which in our case turn out to be $z_1^*$ and $z_2$. What is particularly useful about this approach is that, in this way, we can immediately determine the coefficients of the expansion around the point $z=z_2$ (using \eqref{AO2CP_NC_NE_coeffs}), which is located slightly outside the disk of convergence. After performing this calculation, we, as in the first step, obtain $N_2=12$ coefficients from of the starting $N_1=300$ ones. Again, by using the Puiseux theorem in Eq.~\eqref{f_ser_both}, we have thereby also determined the coefficients $c_n$ of the expansion of the second gapped mode around $z_2$:
\begin{equation}
    \wfr_2(z) = \sum_{n=0}^{N_2-1} c_n (z-z_2)^{n/2},
\end{equation}
with $N_2=12$. The coefficients are compared to their numerically calculated values from the spectral curve in Table~\ref{AdS4_table_of_coefficients_2}.

\begin{table}[ht!]
\centering
\begin{tabular}{|c||c|c||} 
 \hline
 $n$ & $c_n$ (from SC) & $c_n$ (reconstructed) \\ [0.5ex] 
 \hline\hline
0 & $1.6614213311-3.0847448533i$ & $1.6614213311-3.0847448533i$ \\
1 & $0.6041101972+1.0936535299i$ & $0.6041101972+1.0936535299i$ \\
2 & $0.2885621045-0.9107043118i$ & $0.2885621045-0.9107043118i$ \\
3 & $-0.4277129705+0.5774501453i$ & $-0.4277129706+0.5774501454i$ \\
4 & $0.4035764247-0.2298675781i$ & $0.4035764231-0.2298675844i$ \\
5 & $-0.2216353115+0.1017772995i$ & $-0.2216352850+0.1017772729i$ \\
6 & $0.1078634137-0.1688650054i$ & $0.1078639326-0.1688639506i$ \\
7 & $-0.1298281746+0.2281561120i$ & $-0.1298308461+0.2281601714i$ \\
8 & $0.1460876073-0.1885941382i$ & $0.1460284334-0.1886707520i$ \\
9 & $-0.0621106150+0.1583162396i$ & $-0.0619934086+0.1580220644i$ \\
10 & $-0.0314855122-0.2321980090i$ & $-0.0283393657-0.2294904766i$ \\
11 & $0.0316714488+0.3290014052i$ & $0.0297005943+0.3402224777i$ \\
 \hline 
\end{tabular}
\caption{Comparison between the $N_2=12$ coefficients $c_n$ form the series expansion of $\wfr_2(z)$ calculated directly from the spectral curve (SC) and those reconstructed by using the Darboux theorem.
}
\label{AdS4_table_of_coefficients_2}
\end{table}

We can also calculate the value of the gap $\wfr_2(0)$ up to the precision of $1$ significant figure by using the Pad\'e approximant to perform the analytic continuation. The values of the calculated gap $\wfr_2^{\text{calc}}(0)$ and the numerically computed gap $\wfr_2(0)$ from the spectral curve are:
\begin{align}
    \wfr_2^\text{calc}(0)&=2.16275 - 3.25341 i, \\
    \wfr_2(0) &= 2.12981 - 3.28100 i.
\end{align}

Finally, by performing a successive sequence of further analogous steps and by exploring the Riemann surfaces of different modes to locate the level-crossing critical points of the associated spectral curve, we claim that one can obtain the entire spectrum of transverse momentum excitations in this theory.   

\section{Summary and future applications}\label{sec:summary}

In this paper, we addressed the question of when certain types of QFT spectra of correlation functions can be reconstructed from partial knowledge, in particular, here, from the knowledge of the dispersion relation of one of the physical modes in momentum space. While this is an old problem that has appeared in the literature in many incarnations, we believe that this work can serve as an encouragement or even a general statement that far-reaching reconstructions are possible given certain conditions: most importantly, for all modes that are connected via level-crossings at critical points of the associated correlator's spectral curve. Without loss of generality in the proposed procedure, we chose to focus on examples that include a gapless mode of which the dynamics can be investigated through the use of an EFT like hydrodynamics. Beyond making general statements, our goal was to develop a systematic algorithm that could be applied when only a limited amount of information about a single mode was known. To facilitate concrete steps in the reconstruction, we developed an algorithm based on the theorems of Darboux and Puiseux, with the occasional need to perform additional analytic continuations of the dispersion relations within the same Riemann sheet, which can be done in any one of numerous ways. This algorithm should be seen as complementary to the more frequently employed method of Pad\'{e} approximants for convergent series (characterising dispersion relations), used for example in Ref.~\cite{Withers:2018srf}. While it is generally difficult to predict in which situation which method will be more efficient and work `better', what we claim is that re-expansions of series based on the method of Darboux allow for a more precise and rigorous control of the reconstruction procedure as they do not depend on the choice of the order of the Pad\'{e} approximant, the inevitable creation of (a multitude or, as the order grows, potentially `infinitely' many) spurious poles and the need to choose the location of the critical point at the accumulation point of those poles. The re-expansion method is closer to the original spirit of Weierstrass's analytic continuations. For additional details regarding the comparison between the methods of the Darboux theorem and Pad\'{e} approximants, particulary, as applied to the holographic example studied in Section~\ref{sec:M2}, see Appendix~\ref{app3}.

As for the hope that this procedure may be useful for understanding new physics, it is important that the calculation of the (finite) sequence of coefficients $a_n$ (cf.~Eq.~\eqref{w0_ser_z0}), from which the entire procedure stems, be in some way simpler than the calculation of the full correlator in question. This, in fact, is not difficult to imagine. As discussed in Section~\ref{sec:general}, in hydrodynamics, the coefficients $a_n$ can be computed from linear response theory and the application of higher-point Kubo formulae in the (somewhat tricky and non-commuting) limits of $\omega \to 0$ and $q \to 0$. Moreover, one may also directly apply the methods of (effective) kinetic theory to obtain such coefficients, thereby potentially further simplifying the calculations. Beyond hydrodynamics, we believe that the same procedures can also be applied to other derivative expanded EFTs, for example to chiral perturbation theory in quantum chromodynamics. Questions such as whether one can use the techniques of spectrum reconstruction to determine the masses of mesons or hadrons are something that should be investigated in the future. Furthermore, beyond the interest in QFT spectra, one could simply also use calculations of the type performed in this paper to analyse the quasinormal spectrum of black holes or black branes, without any reference to holography. 

Finally, it should be said that in the end, the chosen method of reconstruction, so long as the method works, is unimportant. What is important from the point of view of physics is that numerous, sometimes all modes in a spectrum of a correlator in an interacting QFT are intimately related and that each can contain information about the physics at {\it all} energy scales, vastly beyond their naively perceived regime of applicability. With this work, our attempt was to make this transparent and also provide a rigorous and practical method for such reconstructions that can stretch from the deep infra-red to the extreme ultra-violet. Perhaps, from the point of view of a Wilsonian EFT (with an infinite number of irrelevant terms), we could say that in such spectra, or even theories, the renormalisation group is actually a group with an existing and practically implementable inverse. 

\acknowledgments{The authors would like to thank Borut Bajc, Casey Cartwright, Matthias Kaminski, Hong Liu and Alexander Soloviev for valuable discussions and comments on the draft of the paper. The work was supported by the STFC Ernest Rutherford Fellowship ST/T00388X/1 and the research programme P1-0402 of Slovenian Research Agency (ARRS).}

\appendix
\section{`Critical origin' analogue of the one critical point algorithm}
\label{app:critical_origin}

\subsection{Modifying the ansatz}
\label{app11}

In this appendix, we discuss the necessary small modifications to the reconstruction algorithm that are required when using the Darboux theorem applied to a Puiseux (and not a Taylor) series. In practice, we focus on cases in which the point where the series is given (we choose it to be the origin $z=0$) is a branch point of the same order as the branch point at $z_1$ where the re-expansion takes place. As in the paper, we take the order to be $p=2$. In such cases, the two functions $r(z)$ and $q(z)$ in the ansatz \eqref{f_crit_exp} are no longer analytic in $|z|<|z_1|$. We instead change the ansatz to
\begin{equation}\label{new_ans}
f(z)\sim \left(r^E(z) + z^{-\nu} r^O(z)\right)\left(z-z_1\right)^{-\nu} + q^E(z) + z^{-\nu} q^O(z),
\end{equation}
as $z\rightarrow z_1$, where, now, the functions $r^E$, $r^O$, $q^E$ and $q^O$ are all analytic in the disk $|z|<|z_1|$.

We can write the Puiseux series representation of $f(z)$ as
\begin{align}
f(z) &= \sum_{n=0}^\infty a_n z^{n/2} = \sum_{n=0}^\infty a_{2n}z^n + z^{1/2} \sum_{n=0}^\infty a_{2n+1}z^n.
\end{align}
The theorem of Darboux (matching of two series) then gives
\begin{align}
&a_{2n} \sim \sum_{k=0}^\infty \frac{(-1)^{k-\nu} (\nu-k)_n r^E_k}{z_{*}^{n-k+\nu} n!}, \\
&a_{2n+1} \sim \sum_{k=0}^\infty \frac{(-1)^{k-\nu} (\nu-k)_n r^O_k}{z_{*}^{n-k+\nu} n!}.
\end{align}
This means that in order to determine the branch point position $z_1$ and its order, we can use only the odd or only the even coefficients $a_n$, obtaining the same results. Moreover, to determine the coefficients $r^E$ and $r^O$ of the function $r^E(z) + z^{-\nu} r^O(z)$ controlling the singularity, we must perform the algorithm on even and odd coefficients separately. 

Similarly, to find the coefficients of the expansion for the two functions $q^E$ and $q^O$, we again make use of an auxiliary function $g(z)$ defined exactly as in Eq.~\eqref{g_fun}:
\begin{align}
g(z) &\equiv \left(z-z_1\right)^{\nu} f(z) = \sum_{n=1}^\infty g_n z^n \nn
&\sim r^E(z)+ r^O(z)\sqrt{z} + \left(z-z_1\right)^{\nu} \left(q^E(z)+q^O(z)\sqrt{z}\right),
\end{align}
so that we can again use two versions of Darboux's theorem for the two sets of coefficients
\begin{align}
&g_{2n} \sim \sum_{k=0}^\infty \frac{(-1)^{k-\nu}(\nu-k)_n q^E_k}{z_{*}^{n-k+\nu} n!} ,\\
&g_{2n+1} \sim \sum_{k=0}^\infty \frac{(-1)^{k-\nu}(\nu-k)_n q^O_k}{z_{*}^{n-k+\nu} n!}.
\end{align}

In conclusion, for cases where the origin is a critical point, one uses the same procedures as described in the paper (either with one or more critical points obstructing convergence), but now, separately on the two sets of (even and odd) coefficients.

\subsection{Unwinding the branch point at the origin}
\label{app12}

Another (perhaps more standard) option for dealing with Puiseux series of the form
\begin{equation}
    \sum_{n=0}^\infty a_n z^{n/2}
\end{equation}
is to use the standard algorithms described in the main text but introduce a new variable $w$ that `unwinds' the branch point at the origin instead of changing the ansatz as in \eqref{new_ans}. The new variable is defined as
\begin{equation}
    z=w^2.
\end{equation}
If we use this transformation, the starting coefficients $a_n$ of the expansion in the $w$-plane are the same as the original ones, while the branch point positions and the coefficients of the re-expansion change compared to their values in the $z$-plane. In particular, if the original critical points are at positions $z=z_i$, then the new ones (in the $w$-plane) are at $w=\pm\sqrt{z_i}$. This fact can prove inconvenient when undertaking further re-expansion steps since the number of critical points becomes doubled. Moreover, when determining the coefficients $b_n$ of the expansion 
\begin{equation}
    \sum_{n=0}^\infty b_n (z-z_1)^{n/2}
\end{equation}
around $z_1$, the algorithm \eqref{coeffs_algo_HG_1} gives us the coefficients $\tilde{b}_n$, which are connected to the original coefficients $b_n$ by expressions
\begin{equation}
\begin{aligned}
    \tilde{b}_{2n} &= \sum_{k=\lceil \frac{n}{2} \rceil}^n (2 w_1)^{2k-n} \binom{k}{n-k} b_{2k}, \\
    \tilde{b}_{2n+1} &= \sum_{k=0}^n (2 w_1)^{2k-n+1/2} \binom{k+1/2}{n-k} b_{2k+1}.
\end{aligned}
\end{equation}
After obtaining the coefficients $\tilde{b}_n$ one therefore has to solve the above equations in order to calculate $b_n$.

\section{The re-expansion algorithm with two complex conjugated critical points (Case~\ref{AO2CP_versions_a}) using Gegenbauer polynomials}
\label{app2}

In dealing with the Darboux theorem applied to a case with two closest critical points (at the boundary of the convergence disk), it is sometimes convenient to use the ansatz from Eq.~\eqref{a_asymp_2cp_Gegenbauer}, repeated here for convenience,
\begin{equation}
    f(z) = \left(z-z_1\right)^{-\nu} \left(z-z_2\right)^{-\nu} R(z) + Q(z).
\end{equation}
Instead of the usual Taylor expansion, one now uses the multi-point Taylor expansion of $R(z)$ around the points $z_1$ and $z_2$:
\begin{equation}
    R(z) = \sum_{k=0}^\infty (\alpha_k + z \beta_k) (z-z_1)^k (z-z_2)^k.
\end{equation}
The above expression is now most conveniently thought of in terms of the generating function for the Gegenbauer polynomials $C_n^\sigma(z)$, which are defined as
\begin{equation}
\frac{1}{(1 - 2z s +s^2)^{\sigma}} = \sum_{n=0}^\infty C_n^\sigma(z) s^n.
\end{equation}
This results in the following expression for the asymptotic form of the coefficients $a_n$ written in terms of the coefficients $\alpha_k$ and $\beta_k$:
\begin{equation}
    a_n \sim \frac{1}{R^n} \sum_{k=0}^\infty \left( \alpha_k C_n^{\nu - k}(\cos\theta) + R \beta_k C_{n-1}^{\nu-k}(\cos\theta)\right).
\end{equation}

Using the above form, one can again define the polynomials $X_n^m$ as
\begin{equation}
    \begin{aligned}
        X_n^0 &= a_n, \\
        X_n^{m+1} &= X_n^m - 2\cos\theta\frac{n+\nu-2m-1}{nR} X_{n-1}^m + \frac{(n+2\nu-3m-2)(n-m-1)}{n(n-1)R^2} X_{n-2}^m,
    \end{aligned}
\end{equation}
which, as $n\rightarrow\infty$, scale as
\begin{equation}
    X_n^m \sim n^{\nu-2m-1}.
\end{equation}
Hence, one can again determine the branch point order and position by setting two consecutive polynomials (in $n$) equal to zero.

Similarly, one can use the polynomials $Y_{\ell,n}^m$ defined by
\begin{equation}
    \begin{aligned}
        Y_{\ell,n}^0 &= a_n ,\\
        Y_{\ell,n}^{m+1} &= Y_{\ell,n}^m - 2\cos\theta\frac{n+\nu-2m-\ell-2}{nR} Y_{\ell, n-1}^m \\
        &\phantom{ = }+ \frac{(n+2\nu-3m-2\ell-4)(n-m-1)}{n(n-1)R^2} Y_{\ell, n-2}^m,
    \end{aligned}
\end{equation}
which behave as
\begin{equation}
Y_{\ell,n}^m \sim \frac{1}{R^{n-2m}} \sum_{k=0}^\infty \left[ \alpha_k \left( f_{k,n}^{\ell,m} C_{n-2m+1}^{\nu-k} + g_{k,n}^{\ell,m} C_{n-2m}^{\nu-k}\right) + R \beta_k \left( h_{k,n}^{\ell,m} C_{n-2m}^{\nu-k} + j_{k,n}^{\ell,m} C_{n-2m-1}^{\nu-k}\right)\right],
\end{equation}
with the coefficients $f_{k,n}^{\ell,m}$, $g_{k,n}^{\ell,m}$, $h_{k,n}^{\ell,m}$ and $j_{k,n}^{\ell,m}$ determined by using the recursion relation that holds for the Gegenbauer polynomials:
\begin{equation}
    n C_n^\sigma(z) - 2(n-1+\sigma) z C_{n-1}^\sigma (z) + (n-2 + 2\sigma) C_{n-2}^\sigma(z) = 0.
\end{equation}
As for the form of the re-expansion algorithm used in the main text, the coefficients $f_{k,n}^{\ell,m}$, $g_{k,n}^{\ell,m}$, $h_{k,n}^{\ell,m}$ and $j_{k,n}^{\ell,m}$ equal to zero for $k=\ell+1,\dots,\ell+m$ and so the expression
\begin{equation}
Y_{\ell,n}^m - \frac{1}{R^{n-2m}} \sum_{k=0}^\ell \left[ \alpha_k \left( f_{k,n}^{\ell,m} C_{n-2m+1}^{\nu-k} + g_{k,n}^{\ell,m} C_{n-2m}^{\nu-k}\right) + R \beta_k \left( h_{k,n}^{\ell,m} C_{n-2m}^{\nu-k} + j_{k,n}^{\ell,m} C_{n-2m-1}^{\nu-k}\right)\right],
\end{equation}
scales as $n^{\nu-2m-\ell-2}$ when $n\rightarrow\infty$. Setting it to zero at multiple $n$ allows us to determine the first $\ell$ coefficients $\alpha_k$ and $\beta_k$.

Having obtained the coefficients $\alpha_k$ and $\beta_k$, the coefficients $r_n$ of the series expansion $\sum_{n=0}^\infty r_n(z-z_1)^n$ of $r(z)$ in \eqref{f_asymp_2cp} around $z=z_1$ are then obtained through matching with the expansion of $(z-z_2)^{-\nu} R(z)$ around $z=z_1$. They are given in terms of $\alpha_k$ and $\beta_k$ by
\begin{equation}
r_n = \sum_{k=0}^n \left(2i R\sin\theta\right)^{2k-n-\nu} \left[2iR\sin\theta \binom{k-\nu}{n-k-1} \beta_k + \binom{k-\nu}{n-k} (\alpha_k + z_1 \beta_k)\right],
\label{r_in_terms_of_alpha_beta}
\end{equation}
where we have again used $z_1=Re^{i\theta}$.

In order to find the coefficients of the function $Q(z)$, we define the auxiliary function $g(z)$:
\begin{equation}
g(z) \equiv \left(1-\frac{z}{z_1}\right)^\nu \left(1-\frac{z}{z_2}\right)^\nu f(z) = \sum_{n=0}^\infty g_n z^n = R(z) + \left(1-\frac{z}{z_1}\right)^{\nu} \left(1-\frac{z}{z_2}\right)^{\nu} Q(z),
\end{equation}
where the series coefficients $g_n$ are given by
\begin{equation}
g_n=\sum_{m=0}^\infty \frac{1}{z_1^{n-m}} \binom{n-m-\nu-1}{n-m} {}_2F_1\left(m-n,-\nu,\nu+1+m-n; e^{2i\theta}\right) a_m.
\end{equation}

We now similarly multi-Taylor expand the function $Q(z)$ as
\begin{equation}
Q(z) = \sum_{k=0}^\infty (\gamma_k + z \delta_k) (z-z_1)^k (z-z_2)^k
\end{equation}
and use the above procedure on the coefficients $g_n$ to calculate $\gamma_k$ and $\delta_k$. In order to get the subleading coefficients of $q(z)$ in the expansion \eqref{f_asymp_2cp}, we compare the expansions of $q(z)$ and $Q(z)$ and get
\begin{equation}
q_n = \sum_{k=0}^n (2iR\sin\theta)^{2k-n} \left[ 2iR\sin\theta \binom{k}{n-k-1} \delta_k + \binom{k}{n-k}(\gamma_k + z_1 \delta_k)\right].
\end{equation}
Note that the above expression is different from Eq.~\eqref{r_in_terms_of_alpha_beta} as in that case we used the coefficients $\alpha_k$ and $\beta_k$ to get the coefficients of $r(z)$, leaving outside the factor of $(z-z_1)^{-\nu}$. In the present case, we use the coefficients $\gamma_k$ and $\delta_k$ to calculate the coefficients of $A(z)$ leaving outside both factors $(z-z_1)^{-\nu}$ and $(z-z_1^*)^{-\nu}$.

In this manner, we reconstruct the full Puiseux series representation of $f(z)$ around one of the two closest critical points ($z_1$), where again,
\begin{equation}
f(z) = f_+ (z) =  - i \sum_{n=0}^\infty e^{\frac{i \pi n}{2}} b_n \left(z - z_1\right)^{n/2},
\end{equation}
with $b_n$ given in terms of $r_n$ and $q_m$,
\begin{equation}
b_{2n} = i q_n e^{-in\pi} , \qquad b_{2n+1} = \frac{r_n}{\sqrt{z_1}}e^{-i\left(n+\frac{1}{2}\right)\pi}.
\end{equation}
The two branches of expansions are then related by the following expression:
\begin{equation}
f_\pm (z) =  - i \sum_{n=0}^\infty e^{\pm\frac{ i \pi n}{2}} b_n \left(z - z_1\right)^{n/2}.
\end{equation}

\section{Comparison between the methods based on the Darboux theorem and the Pad\'{e} approximant}
\label{app3}

Perhaps the simplest and often very effective way to address the problem of finding an analytic continuation of a series and subsequently re-expanding it around a branch point is by the method of the Pad\'e approximant. This method was for example used in the analysis of a holographic model in Ref.~\cite{Withers:2018srf}. There, when the Pad\'e approximant was applied to a series such as \eqref{w0_ser_z0}, its poles converged to the closest (convergence-limiting) branch points and the second Riemann sheet could be obtained with the use of the branch point unwinding described in Appendix~\ref{app12} directly from the Pad\'e approximant. Here, we first describe this well-established method and then compare it with the one based on the Darboux theorem for a single step re-expansion in our main example of the holographic theory in the AdS${}_4$-Schwarzschild black brane bulk, which was discussed in Section~\ref{sec:M2}.

Given some function $f(z)$, we define its Pad\'e approximant of order $[M_N,M_D]$ around the point $z=z_0$ as
\begin{equation}
    \mathcal{P}(z;z_0) = \frac{\sum_{n=0}^{M_N} A_n (z-z_0)^n}{1 + \sum_{n=1}^{M_D} B_n (z-z_0)^n},
\end{equation}
where the coefficients $A_n$ and $B_n$ are chosen so that the Taylor expansions around $z=z_0$ of the function $f(z)$ and its approximant $\mathcal{P}(z;z_0)$ match up to order $M_N+M_D$. Given the Taylor expansion representation of a hydrodynamic mode $\wfr_0(z)$ around $z=0$,
\begin{equation}\label{wfr_series}
    \wfr_0(z) = \sum_{n = 1}^{N} a_n z^n,
\end{equation}
a simple choice one can make is to take the Pad\'e approximant $\mathcal{P}_0(z;0)$ of order $[N/2,N/2]$. One may then attempt to use this rational function to first determine the position(s) of the closest branch point(s) to $z = z_0$ in the complex $z$-plane.

\begin{figure}[ht]
    \centering
    \includegraphics{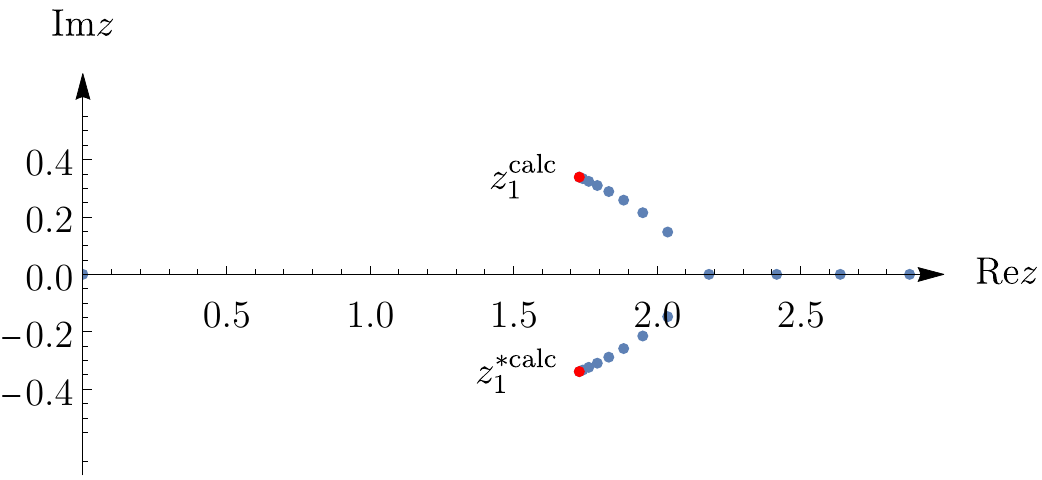}
    \caption{Locations of the poles of the Pad\'e approximant $\mathcal{P}_0(z;0)$ of order $[40,40]$. With red colour, we plot the two poles (`accumulation points') that we use as the approximate locations of the two closest critical points $z_1$ and $z_1^*$.}
    \label{fig:pade_poles}
\end{figure}

As expected, for our example discussed in Section~\ref{sec:M2}, we find that the poles of $\mathcal{P}_0(z;0)$ do indeed accumulate at the locations of the closest pair of branch points $z_1$ and $z_1^*$, `signifying' the closest non-analyticities of the function $\wfr_0(z)$. We show this in Figure~\ref{fig:pade_poles} by plotting the positions of the poles in the $z$-plane.

In order to calculate the coefficients of the series around $z_1$, we must not directly Taylor expand $\mathcal{P}_0(z;0)$ around $z_1$, as this expansion is of course divergent. Instead, we introduce a new variable $u$ (see also Ref.~\cite{Withers:2018srf}):
\begin{equation}
    \frac{z}{z_1} = -u(u+2).
\end{equation}
In this way, we `unwind' the branch point at $z=z_1$ and move the origin of one sheet to $u=0$ while the other is then at $u=-2$. The branch point is now at $u=-1$. Inserting $z=-z_1 u (u+2)$ into the original series \eqref{wfr_series}, i.e., $\wfr_0(-z_1 u (u+2))$, and constructing the Pad\'e approximant around $u=0$ and $u=-2$, we obtain the approximants $\mathcal{P}_0(u;0)$ and $\mathcal{P}_1(u;-2)$ for the two sheets, respectively. This is analogous to the connection between Eqs.~\eqref{w0_ser_z1} and \eqref{w1_ser_z1}. Expanding the approximant $\mathcal{P}_1(u; -2)$ around $u=-1$ then gives us the desired Puiseux series: the first gapped mode's dispersion relation $\wfr_1$ expanded around the leading level-crossing critical point $z_1$ where $\wfr_1$ collides with $\wfr_0$.

We now also briefly comment on the obtained results. By using the same number of starting coefficients $N_0=300$ as in Section~\ref{sec:M2}, this procedure determines the closest branch point $z_1^\text{calc}$ to only $3$ significant figures. This is rather poor compared to the $18$ significant figures precision that we obtained from the Darboux theorem (without any optimisation). Therefore, when it comes to the determination of the branch point position in this example, the method of the Pad\'{e} approximant clearly underperforms. With the calculated $z_1^\text{calc}$, the precision of the coefficients that follow from $\mathcal{P}_1(u;-2)$ is also expectedly poor. In particular, we satisfactorily determine only the first coefficient of the Puiseux series expansion around $z_1$ to $1$ significant figure precision, while all higher coefficients are completely unreliable. Nevertheless, what is fascinating is that even with the branch point determined to only $3$ significant figures, one is still able to obtain the value of the gap by directly evaluating the Pad\'e approximant $\mathcal{P}_1(u;-2)$ at $u=0$ to the incredible precision of $17$ significant figures. This has to be compared with the precision of $2$ significant figures with which we calculated the gap using a combination of the Darboux theorem and the Pad\'{e} approximant (as the analytic continuation within the same Riemann sheet).

To further test and compare the methods, it is also instructive to use the numerically obtained value of the branch point $z_1$ that can be determined directly from the spectral curve. The results that follow from the Pad\'e approximant are then dramatically better than those obtained from the Darboux theorem. One can determine the gap to the precision of $26$ significant figures and calculate the first $80$ coefficients of the Puiseux series to precision of more than $10$ significant figures.

In conclusion, we find that for the holographic example we studied in this work, the Pad\'e approximant method is less effective than the Darboux theorem. The main reason for this seems to be the error that arises from determining the location of the relevant branch point. However, if the branch point is for some reason known exactly, then the Pad\'{e} approximant works extremely well. From a different standpoint, we note that the `ad-hoc' way in which the accumulation point of the poles must be selected is arguably the most theoretically unsatisfying feature of Pad\'{e} approximants. Moreover, the locations of the spurious poles from the denominator may also cause considerable problems at different stages of the reconstruction. In this sense, the method of Darboux clearly appears to be better controlled. On the other hand, for certain questions, Pad\'{e} approximants are simpler to implement, we saw that they performed extremely well in determining the gap (the approximant is an analytic continuation to certain regions outside the convergence disk in its own right) and can even be used when dealing with asymptotic series. Using a combination of different methods in parallel (which also allows for various cross-checks) therefore unsurprisingly appears to be the best strategy.

\bibliographystyle{JHEP}
\bibliography{Genbib}{}

\end{document}